\documentclass[aps,pre, twocolumn, superscriptaddress, groupedaddress, floatfix]{revtex4}

\usepackage{amsmath}
\usepackage{graphicx}
\usepackage{subfig}
\usepackage{color}
\usepackage[dvipsnames]{xcolor}
\usepackage{epstopdf}

\usepackage{amssymb, amsmath, amsfonts}
\usepackage{multirow,bigdelim}
\usepackage{graphicx}
\usepackage[latin1]{inputenc}
\usepackage{rotating}

\usepackage{array}
\newcolumntype{e}{@{\extracolsep{\fill}}c}

\newcommand{\bl}{_\mathrm{bl}}
\newcommand{\unbl}{_\mathrm{unbl}}
\newcommand{\trans}{^\mathrm{trans}}

\begin{document}

\title{Two-species TASEP model: from a simple description to intermittency and travelling traffic jams}

\author{Pierre Bonnin}

\affiliation{Institute for Complex Systems and Mathematical Biology, Department of Physics, Aberdeen AB24 3UE, UK}
\affiliation{Institute of Medical Sciences, School of Medicine, Medical Sciences and Nutrition, University of Aberdeen, Aberdeen, AB25 2ZD, UK}

\author{Ian Stansfield}
\affiliation{Institute of Medical Sciences, School of Medicine, Medical Sciences and Nutrition, University of Aberdeen, Aberdeen, AB25 2ZD, UK}

\author{M. Carmen Romano}
\affiliation{Institute for Complex Systems and Mathematical Biology, Department of Physics, Aberdeen AB24 3UE, UK}
\affiliation{Institute of Medical Sciences, School of Medicine, Medical Sciences and Nutrition, University of Aberdeen, Aberdeen, AB25 2ZD, UK}

\author{Norbert Kern}
\affiliation{Laboratoire Charles Coulomb (L2C), University of Montpellier, CNRS, Montpellier, France}

\date{\today}

\begin{abstract}
 
We extend the paradigmatic and versatile TASEP (Totally Asymmetric Simple Exclusion Process) for stochastic 1d transport to allow for two different particle species, each having specific entry and exit rates. We offer a complete mean-field analysis, including a phase diagram, by mapping this model onto an effective one-species TASEP. Stochastic simulations confirm the results, but indicate deviations when the particle species have very different exit rates. We illustrate that this is due to a phenomenon of intermittency, and formulate a refined 'intermittent' mean-field (iMF) theory for this regime. We discuss how non-stationary effects may further enrich the phenomenology.
\end{abstract}

\maketitle

\section{Introduction}
Transport processes are ubiquitous in nature and technology. Modelling them mathematically aims at describing and predicting flow of the entity of interest, as well as providing insight into key mechanisms of the process. Different approaches in physics and mathematics have been applied, according to the type of transport phenomena. The case of vehicular traffic is a good example of a phenomenon where a broad range of such mathematical models have been applied. They range from macroscopic descriptions, where traffic is described as a compressible fluid~\cite{Lighthill:1955,Richards:1956}, to microscopic approaches, where the movement of individual vehicles is described in terms of interacting particles far from equilibrium ~\cite{Pipes,Chandler1,Chandler2, Hoogendoorn, Wilson1,Wilson2}. 
Also, statistical physics approaches have been applied to describe vehicular traffic. They have put emphasis on describing the fundamental, general features, by developing models that only incorporate the most essential features. This allows one to unveil the key mechanisms, and thus deepen our understanding~\cite{chowdhury_traffic}.

The Totally Asymmetric Simple Exclusion Process (TASEP) is such a model, which has become a paradigmatic process for studying directed stochastic transport in constrained, quasi-1d geometries subject to excluded volume interactions. One may argue that the force of the model lies in its simplicity, which allows it to shed light generically, with implications for many different transport processes. At the same time, it has been successfully adapted to account for the complexity of many specific transport situations. In this paper we study an extension of the TASEP, exploring the additional features which arise when different kinds of transported particles are discriminated by the rates with which they enter, and then ultimately leave the system.

The TASEP model has been studied extensively in the literature. It has become a key model of non-equilibrium statistical physics, exhibiting a rich phenomenology, such as boundary induced phase transitions~\cite{krug} and shock waves~\cite{derrida_SW}, to cite but two examples. In fact, the TASEP was originally introduced to describe the process of protein synthesis~\cite{MacDonald:68}, and it is still the basis of a large number of models of translation of mRNA into proteins~\cite{shaw, chowdhury, romano, tuller}. At the same time it has found applications in many other fields, such as transcription~\cite{Klumpp18159, lefranc},  intracellular transport of molecular motors~\cite{kern-parmeggiani}, molecular transport across membrane channels~\cite{chou, kolomeisky}, fungal growth~\cite{evans} as well as vehicular and pedestrian traffic~\cite{chowdhury_traffic,appert_pedestrian}. 

In the majority of considered models all particles behave identically, i.e. they all share the same microscopic rates at which the enter the system, hop from one site to the next, and finally exit the system.
In some of these setups, however, it is more realistic to distinguish different types of particles travelling through the lattice. One straightforward example is vehicular transport, where clearly motorbikes, cars and lorries would be expected to enter a main road with very different dynamics, would travel at different speeds, and might also differ in the process by which they exit onto a side road. In pedestrian traffic, different age groups may be described. Similar considerations are expected to hold, on a microscopic scale, for molecular motors: for example, the rates governing their dynamics may vary between different types of motors advancing along microtubules~\cite{alberts}. On a yet smaller scale, in mRNA translation, it is known that two  types of ribosomes can be distinguished, the dynamics of which differ according to whether they have bound certain protein complexes (RAC/NAC complexes) that assist in the process of polypeptide folding \cite{racnac}.

Multi-species TASEP models have been  considered previously. Most such models have been constructed with 'classes' of particles in mind, which do not simply differ in their dynamics but also weaken the excluded volume interactions. In these models, particles pertaining to a class of a higher 'rank' in the class hierarchy are allowed to 'overtake' those of a lower rank, and  it is this distinction which leads to different particle dynamics~\cite{crampe,arita-mallick,ayyer}. These systems have great fundamental interest, as they lead to a rich stochastic process; tracing a small number of such particles of a different class has furthermore proven a useful approach for dynamically locating the edges of high and low density zones in a TASEP transport process~\cite{derrida:1997}.

A more direct distinction in terms of microscopic dynamics has been studied in~\cite{bottero}, where two different types of particles share a one-dimensional lattice on which they advance, while overtaking is not permitted. This model has been introduced to describe the traffic of different types of molecular motors along microtubules. In this work, the two types of particles are considered to differ in their entry and bulk hopping rates, which complexifies the transport dynamics with respect to a single-species TASEP model. Their exit rates, however, were assumed to be identical.

In this paper we focus on the opposite, complementary scenario. We take the bulk hopping rates of all particles to be the same, but distinguish two particle species through particle-specific entry and exit rates. Although this does not complexify the bulk dynamics, we show that in particular the specificity in exit rates is a fundamentally new ingredient, which leads to rich behaviour. We first elaborate a full mean-field description for this system, by mapping it onto an effective simple-species model. We construct a comprehensive phase diagram accounting for various scenarios, according to the values of the four input and exit rates. We show, based on stochastic simulations, that this description correctly reproduces simulation results as long as entry and exit rates are of the same order of magnitude.

In the second part of the paper, we analyse the limiting case where the exit rates differ greatly between particle species. We show that this regime can lead to intermittent dynamics, for which the mean-field approach fails to predict both the particle current and the profile of particle density along the lattice. We then introduce a modified mean-field approach for this intermittent dynamics, and show that it yields a good match to numerical simulations when intermittency is present. We end the paper by discussing the results and the scope of the proposed approach, pointing out further interesting features which the model exhibits in the intermittent regime, as perspectives for further studies.

\section{Model and approach}~\label{sec:model}

To represent transport of two different types of objects along a one-dimensional track, with two sub-populations of particles, we build on the Totally Asymmetric Simple Exclusion Process  model. The standard, single-species TASEP consists of a one-dimensional lattice of $L$ sites, along which particles of a single type are transported \cite{Derrida:1992}. Particles attempt to enter the lattice at site $i=1$ with rate $\alpha$ and they leave the lattice at site $i=L$ with rate $\beta$. At the bulk sites ($i=2, \ldots,L-1$) particles hop stochastically from site $i$ to site $i+1$ with rate $\gamma$, provided that site $i+1$ is not occupied.

One way to summarise the key features of the TASEP in a condensed way is by thinking in terms of which process limits the flow. The hopping process in the bulk sets an upper limit to the current. Indeed, a simple mean-field argument suggests a bulk current of $\gamma \, \rho \, (1-\rho)$, where $\rho$ is the bulk density. The maximum current (MC) phase therefore corresponds to a current of $J_{MC}=\gamma/4$, achieved at a bulk density of $\rho_{MC}=1/2$, whenever the limiting rate is the bulk hopping rate. In contrast, when particles enter at a small rate, then this process limits the current. In this case, a low density (LD) phase arises with a density set as $\rho_{LD}=\alpha/\gamma$, with a corresponding current of $J_{LD}=\alpha \, (1-\alpha/\gamma)$. The opposite case arises when the exit rate limits the transport. In that case, we are dealing with a high density (HD) phase, for which a bulk density of $\rho_{HD}=1-\beta/\gamma$ leads to a current of $J_{HD}=\beta \, (1-\beta/\gamma)$. The beauty of the TASEP model is underpinned by two observations. First, this straightforward analysis is indeed key to understanding the transport features, or at least so once a 'phase diagram' is established, which we return to in the following. Second, somewhat surprisingly, the simplified mean-field arguments sketched above turn out to reproduce the {\it exact} results in the limit of an infinite lattice ($L \to \infty$), as has been show by a variety of arguments~\cite{Schutz:exact,Derrida:exact,Evans:exact}.

\begin{figure}
  \centering
  \includegraphics[width=0.5\textwidth]{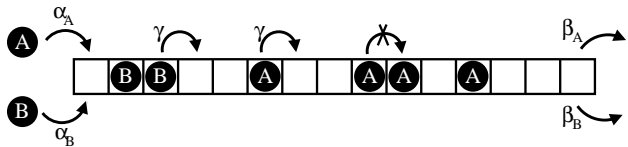}
  \caption{  \label{TASEP2}
       Sketch illustrating the model. Particles are injected onto the first site with an entry rate depending on the species ($\alpha_A$ and $\alpha_B$, respectively). Particles stochastically advance one site at a time, with rate $\gamma$, subject only to the next site being free. The 'bulk' hopping rate $\gamma$ is the same for both types of particles. The exit rates ($\beta_A$ and $\beta_B$) on the last site are again specific to the species.}
\end{figure}

Here we study an extension of the TASEP model.  We consider two categories of particles, which we label $A$ and $B$, to which a given particle belongs for its  entire journey along the segment. Both types of particle step along the lattice stochastically, at the same rate $\gamma$, according to the exclusion process. Thus particles have excluded volume interactions, implying they can neither occupy the same site nor overtake one another. However, particle species are distinguished by their entry rates ($\alpha_A$ and $\alpha_B$) as well as their exit rates ($\beta_A$ and $\beta_B$).
This is illustrated schematically in terms of a two-population TASEP model in Fig.\ref{TASEP2}. The model may alternatively be viewed as a non-markovian single-species TASEP, in which the waiting time at the exit site is drawn from two different exponential distributions, the choice of which has been attributed to each particle as it enters the system. As the notion of particle species is natural for the biological situation of mRNA translation, we will adopt the language of the two-species model in the following.
\\

\section{Two-species TASEP: mean-field approach \label{sec:mfdescription}}

Essentially, we are dealing with a TASEP in which two different species compete on a given segment. As both hopping rates are identical, we can thus think of an effective single-species TASEP, for which the mean-field approach makes it possible to establish the corresponding effective entry/exit rates. We follow up this simple approach, showing that it captures the process in many cases, before addressing its failure when  in/out-rates differ greatly between species.

\subsection{Effective entry and exit rates}
We take the bulk hopping rate, which is the same to both species, to be equal to one ($\gamma = 1$). Both species contribute indifferently to the bulk dynamics. The total density of particles $\rho$ is therefore
\begin{equation} \label{eq:totaldensity} 
  \rho = \rho_A +\rho_B
  \ ,
\end{equation}
which is a sum of the partial densities of each species, {\it i.e.} $\rho_A = N_A/L$ for $N_A$ particles of type A on a lattice of length $L$, and similar for particles of type B. The total current can be written as
\begin{equation} \label{eq:fundamentals2.2}
  J = J_A  + J_B
\end{equation}
in a similar fashion. 
It is important to note that, since particles do not drop off the lattice, currents for both populations are preserved along the segment, from the first to the last site:
\begin{equation}
 \label{eq:AllCurrents}
  J_{A}^{(0)}=J_{A}^{(L)} = J_A
  \ ,
  \quad \mbox{and} \quad
  J_{B}^{(0)}=J_{B}^{(L)} = J_B
  \ .
\end{equation}
Here $J_{A,B}$ denote the bulk currents (crossing any site on the lattice), which are thus also equal to their corresponding currents  entering the segment $J_{A,B}^{(0)}$ and to their corresponding currents leaving the segment $J_{A,B}^{(L)}$. 
\\

At any given site, the mean-field expression for either of the partial currents is the product of two probabilities: the probability of having a particle of the considered species and the probability of having an empty site ahead:
\begin{equation}
  \label{eq:PartialCurrents}
  J_{A,B} = \rho_{A,B} \, (1-\rho)
  \ ,
\end{equation}
where we have taken the density profile $\rho_i$ to be flat, as is known to be justified in the bulk region~\cite{krug}. By writing this relation, as well as the following ones, we make the same mean-field hypothesis as in the standard TASEP model, i.e. we neglect correlations in the occupancy of successive sites.

The expression (\ref{eq:PartialCurrents}) for the partial currents  directly implies that, irrespective of what happens at the boundaries, their ratio is simply identical to that of the partial densities: $\frac{J_A}{J_B}=\frac{\rho_A}{\rho_B}$.
Summing the partial currents furthermore shows that the total current obeys the classical expression for TASEP:
\begin{equation}
  \label{eq:totalcurrent}
  J = J_A + J_B
  = \rho \, (1-\rho)
  \ .
\end{equation}

Now, equating the expressions for the partial in-currents and out-currents, 
\begin{eqnarray}
  \label{eq:InOutCurrents}
  J_{A,B}^{(0)} &= \alpha_{A,B} \, (1-\rho^{(1)})
  \;\mbox{and}\;
  J_{A,B}^{(L)} &= \beta_{A,B} \, \rho_{A,B}^{(L)}
\end{eqnarray}
leads to the observation that the ratio of partial densities $\rho_{A,B}^{(L)}$ at the exit site is directly set by the partial entry/exit rates $\alpha_A$ and $\alpha_B$:
\begin{equation} \label{eq:densityfraction}
  \frac{\alpha_A}{\alpha_B}
  =
  \frac{\beta_A \, \rho_{A}^{(L)}}{\beta_B \, \rho_{B}^{(L)}}
  \ .
\end{equation}
This relation will prove central in the following.
\\

We now turn to establishing the effective entry/exit rates. Writing the total current at the first site as		
\begin{equation}
  J^{(0)}
  = J_A^{(0)} + J_B^{(0)}
  = (\alpha_A+\alpha_B)(1-\rho^{(1)}) \,
\end{equation}
directly arises in the form of an inflowing current, $J^{(0)} = \alpha_{eff} \, (1-\rho^{(1)})$, where the effective entry rate $\alpha_{eff}$ is thus given as
\begin{equation} \label{eq:effective:alpha} 
  \alpha_{eff} = \alpha_A + \alpha_B
  \ .
\end{equation}

To derive an expression for the effective exit rate $\beta_{eff}$ we start from Eq.~\ref{eq:densityfraction} and obtain:
\begin{equation} \label{eq:effective:ratio} 
  \begin{split}
    \frac{\rho_{A}^{(L)}}{\rho_{B}^{(L)}} &= \frac{\beta_B \, \alpha_A}{\beta_A \, \alpha_B}
    \ .
  \end{split}
\end{equation}
	
Eliminating $\rho_A^{(L)}$ via Eq.~(\ref{eq:totaldensity}) yields an en explicit expression for the density of $B$ particles at the exit site $\rho_B^{(L)}$:
\begin{equation}
    \rho_{B}^{(L)}
    = \frac{\beta_A \, \alpha_B}{\beta_B\alpha_A+\beta_A\alpha_B} \, \rho^{(L)}
    \ .
\end{equation}
Analogously, we have%
\begin{equation}
  \begin{split}
    \rho_A^{(L)} &= \frac{\beta_B \, \alpha_A}{\beta_B\alpha_A+\beta_A\alpha_B} \,  \rho^{(L)}  \, ,
  \end{split}	
\end{equation}
as is seen either from applying Eq. (\ref{eq:totaldensity}) to the last lattice site, or simply from symmetry permuting indices A and B.

The exit current can now be written, from Eq. (\ref{eq:totalcurrent}) evaluated at the last site,
\begin{equation}
    J^{(L)}
    = \beta_A \, \rho_A^{(L)} + \beta_B \, \rho_B^{(L)}\\
    =  (\alpha_A+\alpha_B) \, \frac{\beta_A \, \beta_B}{\beta_B \, \alpha_A+\beta_A \, \alpha_B}  \, \rho^{(L)} 
    \,
\end{equation}
By doing so, we assume that the mean-field hypothesis of uncorrelated site occupancies remains valid, despite the new feature introduced by the species-dependent distribution of waiting times.
Again, by analogy to the TASEP mean-field expression, the effective exit rate is thus
\begin{equation} \label{eq:effective:beta} 
  \beta_{eff} = (\alpha_A+\alpha_B) \, \frac{\beta_A\beta_B}{\beta_B\alpha_A+\beta_A\alpha_B}.
\end{equation}

Note that Eqs. (\ref{eq:effective:alpha}) and (\ref{eq:effective:beta}) imply a simple relation for the ratio between the effective rates:
\begin{equation}
  \label{effectiverates:ratio}
  \frac{\alpha_{eff}}{\beta_{eff}} = \frac{\alpha_A}{\beta_A} + \frac{\alpha_B}{\beta_B}
  \ .
\end{equation}

It will also be useful to underline that the relations setting the effective rates, Eqs.~\ref{eq:effective:alpha} and \ref{eq:effective:beta}, have direct physical significance.
Indeed, the abundance of each type of particles can be characterised by the partial densities $\chi_A$ and $\chi_B=1-\chi_A$, defined as
\begin{equation}
  \label{eq:chiAB:def}
  \chi_A = \frac{\rho_A}{\rho_A+\rho_B}
  \qquad\mbox{and}\qquad 
  \chi_B = \frac{\rho_B}{\rho_A+\rho_B}
  \ .
\end{equation}

  Using successively Eqs.~(\ref{eq:PartialCurrents}) and (\ref{eq:AllCurrents}), as well as current conservation, these can be expressed as
  \begin{equation}
  \chi_{A}
  = \frac{J_A}{J_A+J_B}
  = \frac{J_A^{(0)}}{J_A^{(0)}+J_B^{(0)}}
  \ ,
  \end{equation}
  and similar for $\chi_B$. From Eq.~(\ref{eq:InOutCurrents}) we then have  
\begin{equation}
  \label{eq:chiAB}
  \chi_A = \frac{\alpha_A}{\alpha_A+\alpha_B} \in [0,1]
  \;\mbox{and}\; 
  \chi_B = \frac{\alpha_B}{\alpha_A+\alpha_B} \in [0,1],
  \,
\end{equation}
and therefore the partial density of each species is set directly by the percentage with which it contributes to the total input rate, as it is of course expected.
\\

Regarding the effective exit rate, we can now re-write  Eq. (\ref{effectiverates:ratio}) by dividing out $\alpha_{eff}=\alpha_A+\alpha_B$ to obtain
\begin{equation}
  \label{eq:betaeffinverse}
  \frac{1}{\beta_{eff}} = \chi_A \, \frac{1}{\beta_A} + \chi_B \, \frac{1}{\beta_B}
  \ .
\end{equation}
Since an inverse exit rate corresponds to the average time required for a particle of a given type to exit from the last site, the effective exit rate thus corresponds to the population-weighted average of these exit rates.

Note that all results stated so far are valid for any choice of parameters, to the extent that the mean-field approach holds. We will first explore this mean-field behaviour, and then show how it breaks down under specific conditions.

\subsection{Mean-field phase diagram}

Based on the expressions for the effective rates in the mean-field approximation we can now establish the phase diagram for the model with two types of particles.
For any given set of entry/exit rates, the mean-field behaviour of the model is therefore characterised by mapping it onto a corresponding standard single-species model with the appropriate effective rates. However, changing any of the entry/exit rates modifies this mapping, and therefore may potentially drive the system across a phase boundary. Ideally one would like to be able to establish how modifying a single rate, or modifying several rates simultaneously, affects the phases to be observed in the system. This is what we set out to do here.

\begin{figure*}
  \parbox{\textwidth}{
  \hfill
  \subfloat[]{
    \label{fig:tasep:phases}
    \includegraphics[width=0.3\textwidth]{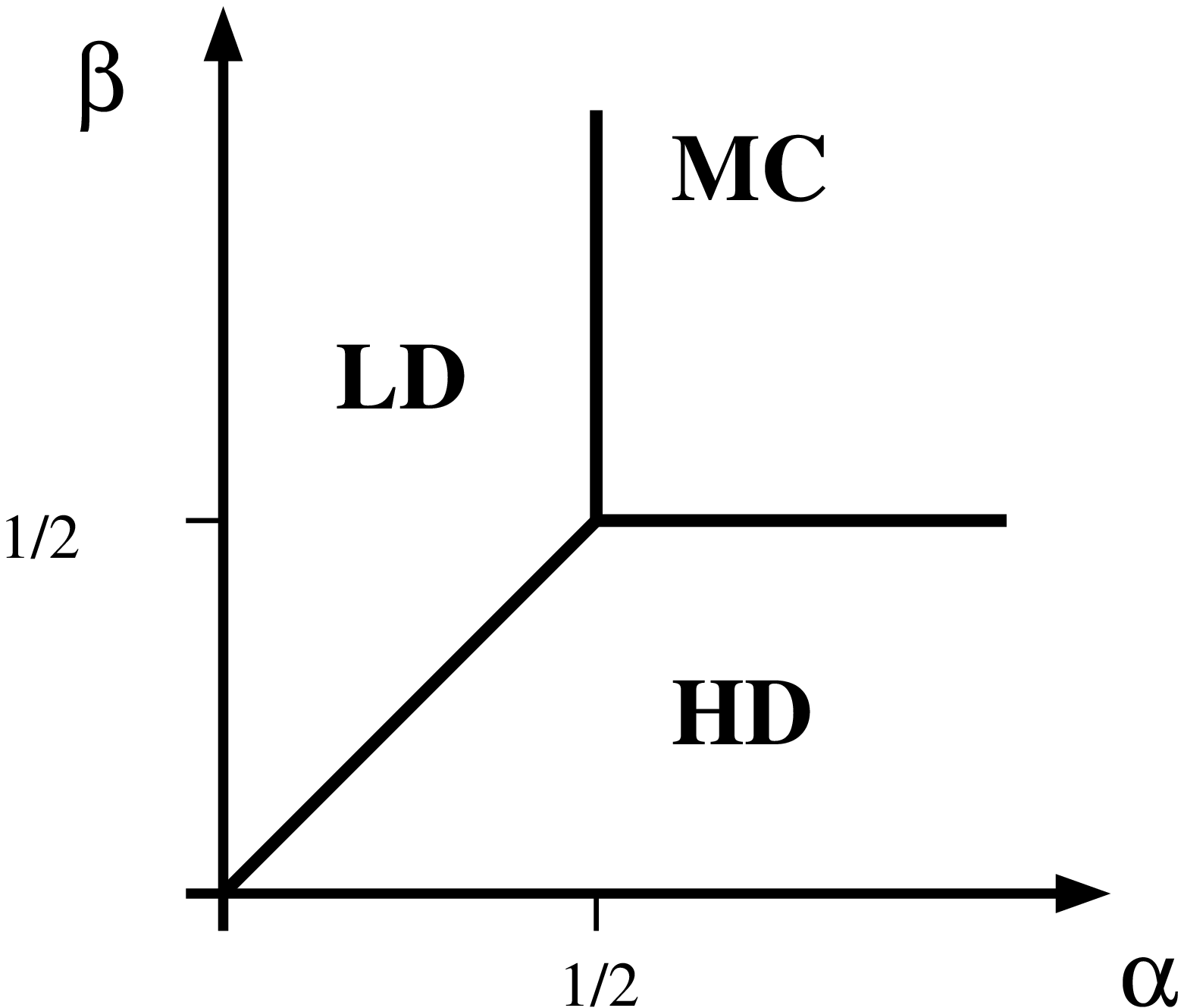}
  }
  \hfill
  \subfloat[]{
    \label{fig:tasep:conditions}
    \includegraphics[width=0.3\textwidth]{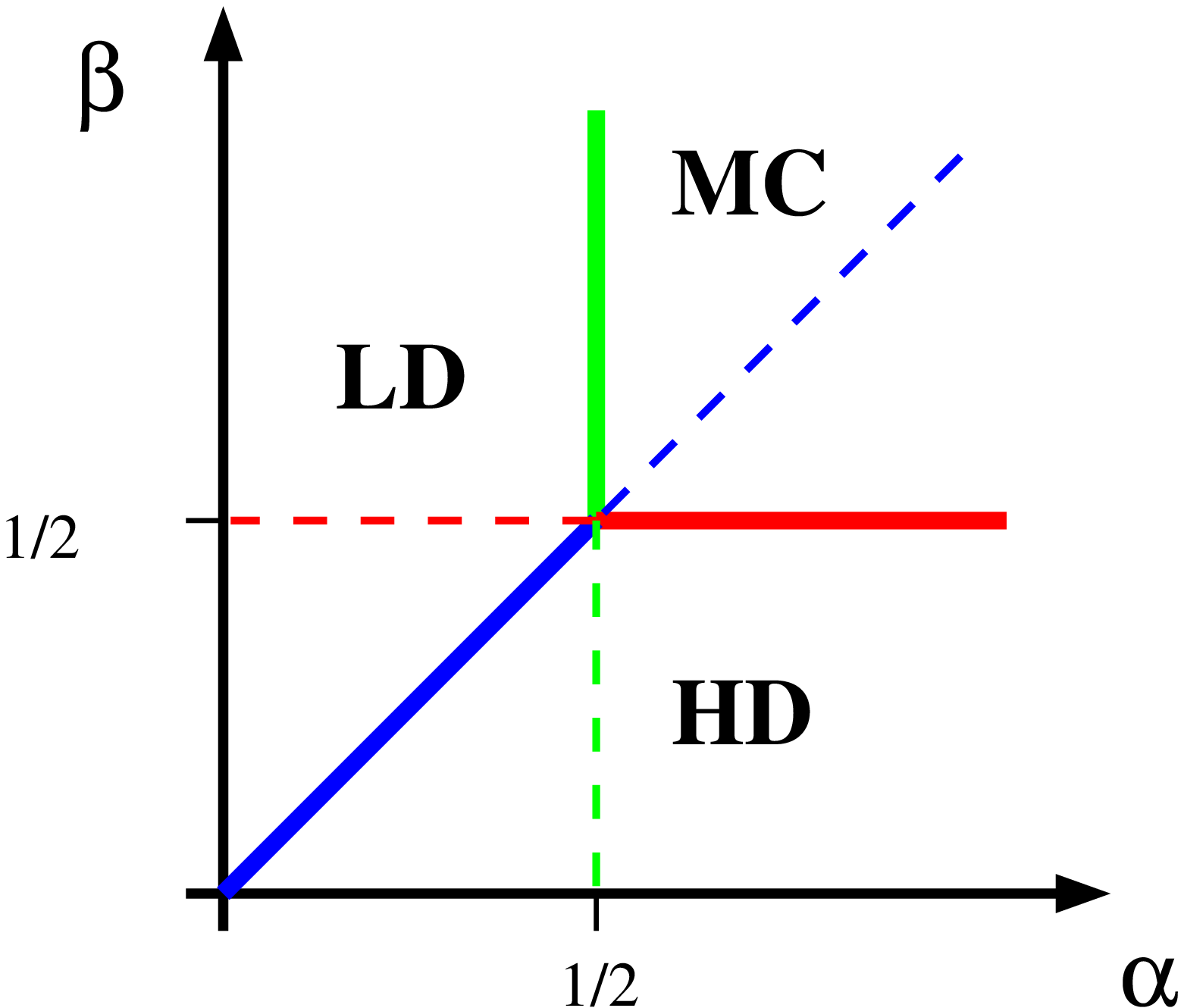}
    }
  \hfill\hfill
  }
  \\
  \hfill
  \subfloat[]{
    \hfill
    \parbox[c]{\textwidth}{
      \parbox[c]{4.5cm}{
        \includegraphics[width=0.2\textwidth]{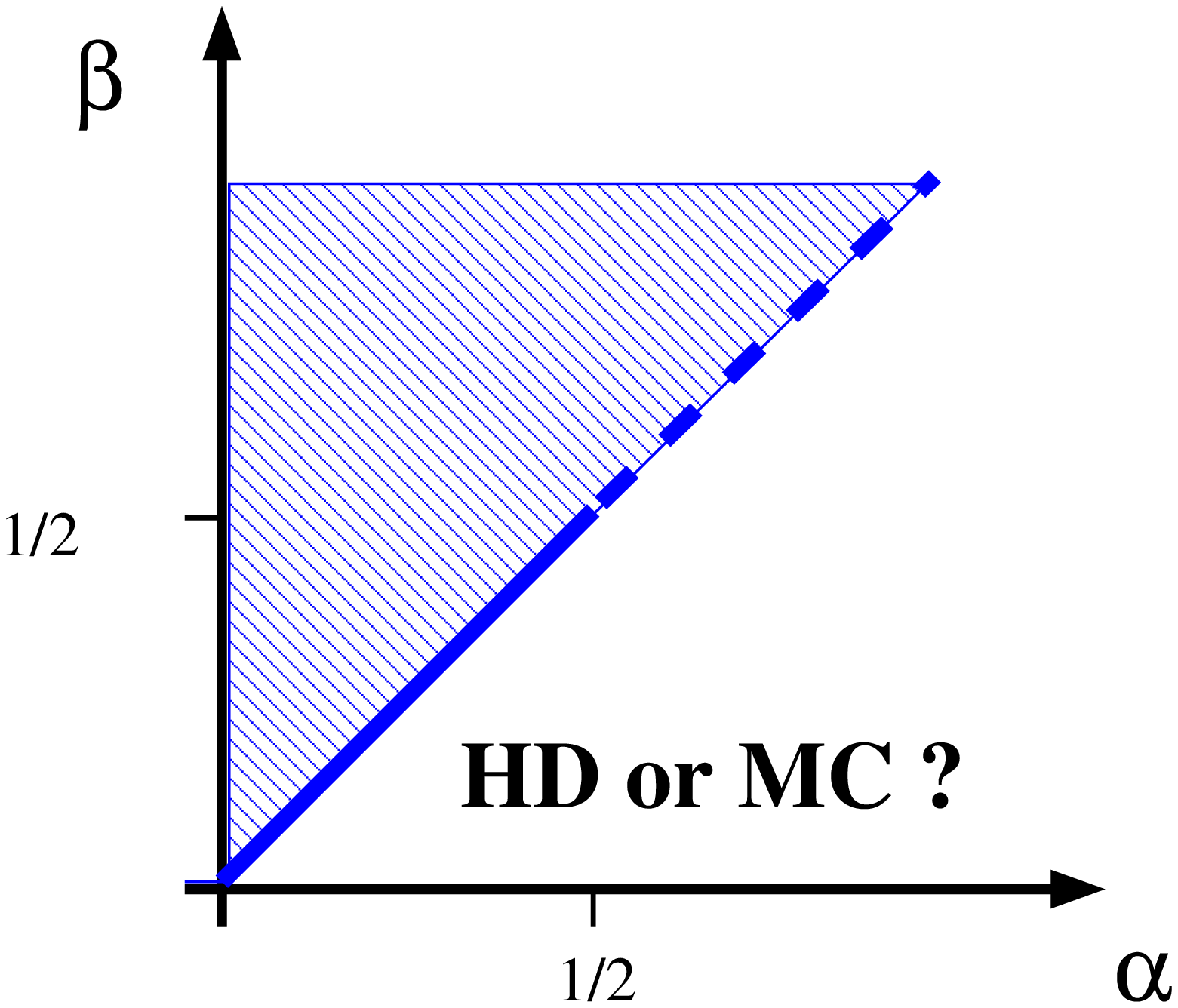}
       }
\hfill
 \parbox[c]{4.5cm}{
	\includegraphics[width=0.2\textwidth]{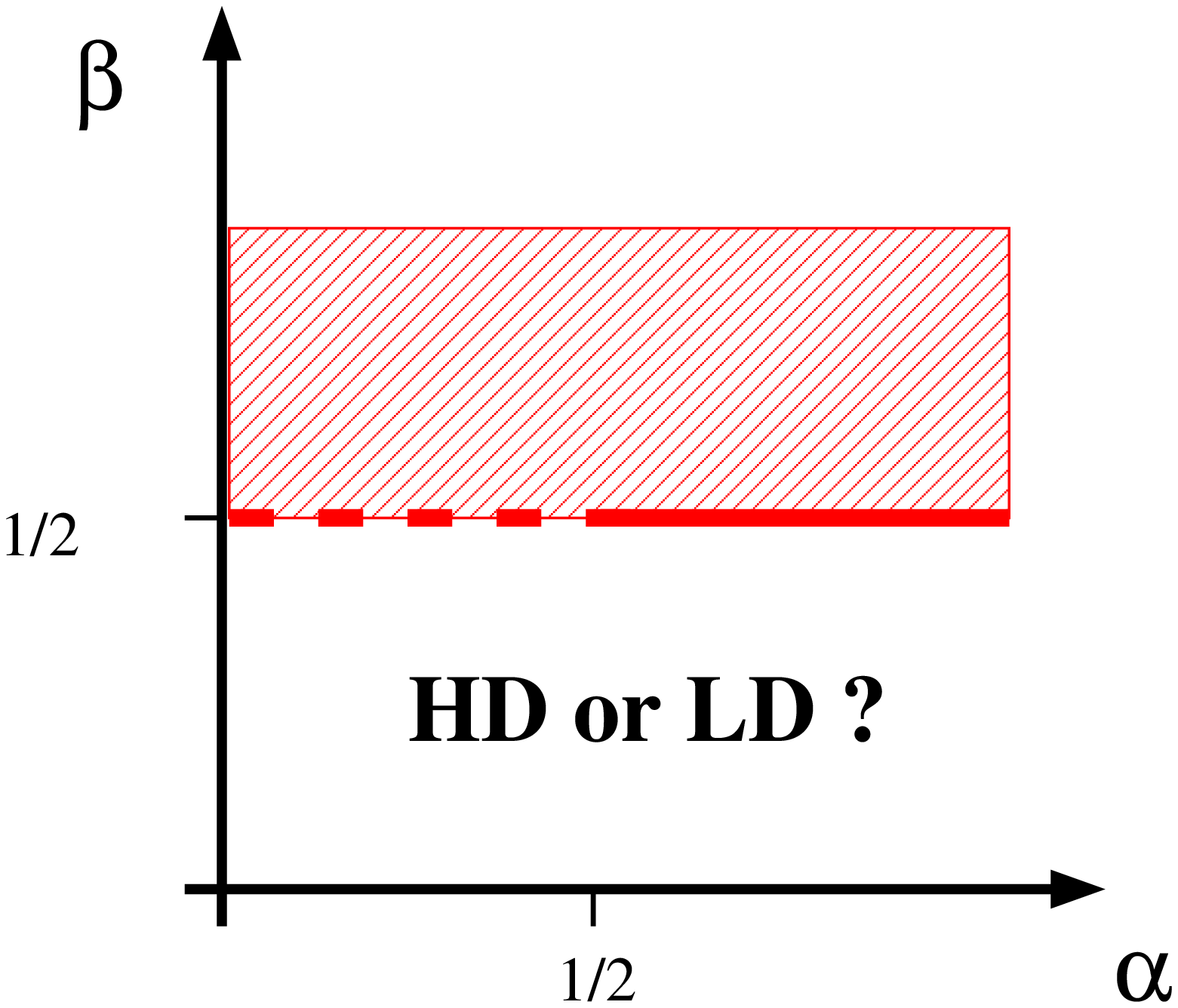}
      }
 \hfill
 \parbox[c]{4.5cm}{        
   	\includegraphics[width=0.2\textwidth]{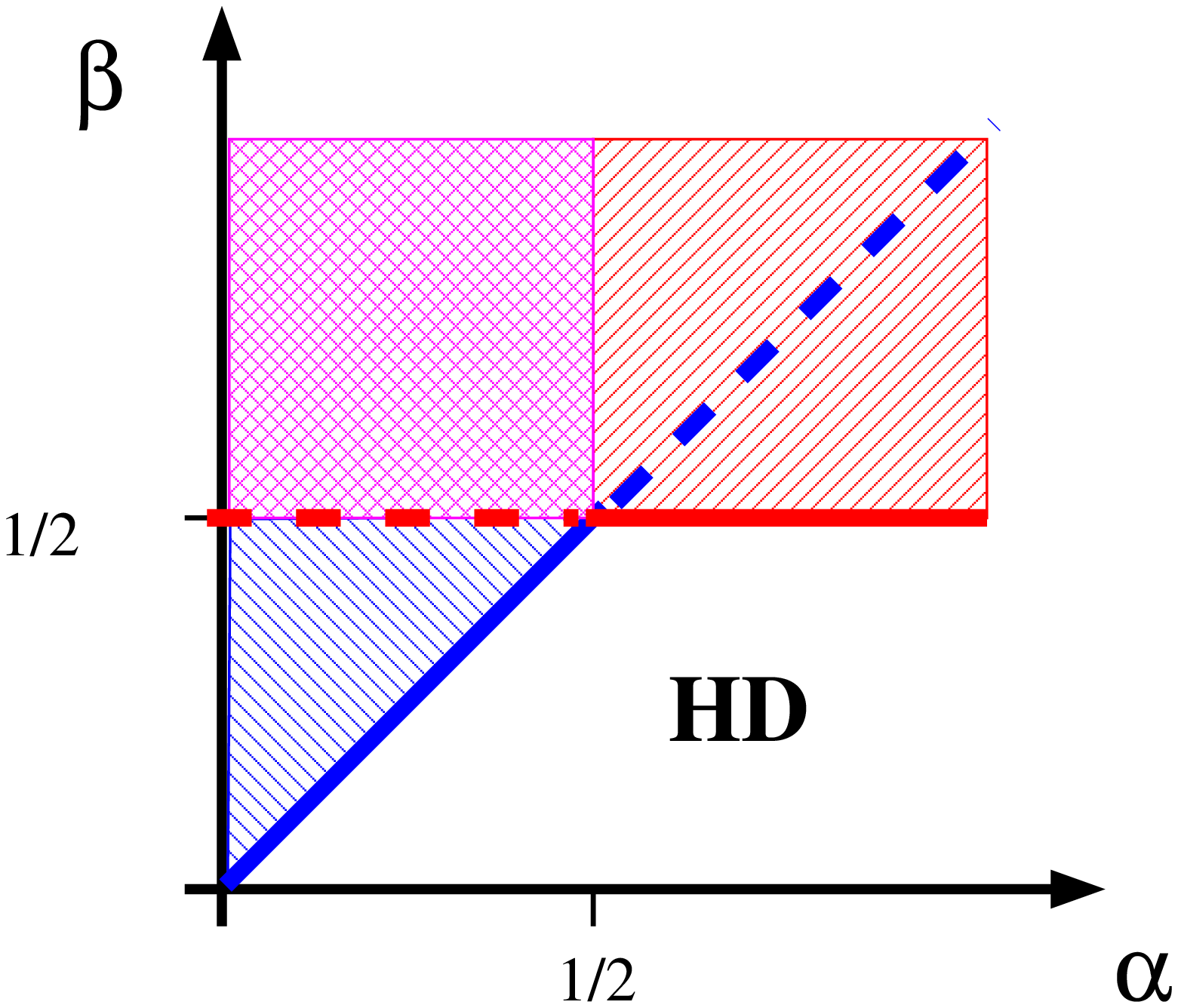}
      }
    }
    \hfill
  }
  \caption{  \label{fig:TASEP:prevailing}
    Single-species TASEP. The phase diagram (panel (a)) is constructed based on three conditions (panel (b)): $\alpha=\beta$ (blue) at the boundary between LD and HD, $\alpha=1/2$ (green) at the boundary between LD and MC, and $\beta=1/2$ (red) between HD and MC. The dashed lines extrapolate these conditions into regions where they do not discriminate phases. The same colour code is preserved later, in order to clarify how the two-species phase diagram can be constructed.
    (c) Phases can be determined by cumulating two conditions, with the common ground of them identifying the corresponding zone in the phase plane. The same argument is used below for the model with two species.
  }
\end{figure*}

We label the phases, just as in the standard, single-species TASEP, as LD, HD or MD, according to whether their (total) density in the bulk is inferior, superior or equal to 1/2. The conditions fixing the well-established single-species phase diagram, summarised in Fig. \ref{fig:tasep:phases}, are:
\begin{equation}
  \label{eq:tasep:conditions}
  \begin{array}{lllllllll}
    \mbox{(LD) } & \mbox{(i) } \ \alpha<\beta  & \mbox{ and } & \mbox{(ii) } \ \alpha<1/2
    \\
    \mbox{(HD) } & \mbox{(i) } \ \beta<\alpha  & \mbox{ and } & \mbox{(ii) } \ \beta<1/2
    \\
    \mbox{(MC) } & \mbox{(i) } \ \alpha>1/2    & \mbox{ and } & \mbox{(ii) } \ \beta>1/2
  \end{array}
\end{equation}

We thus need to exploit these criteria in terms of the effective rates for the two-species model, Eqs. (\ref{eq:effective:alpha}) and  (\ref{eq:effective:beta}). As these depend on all four in/out-rates, it might appear necessary to consider many cases separately, with different scenarios for the phase diagram. However, the discussion can be largely simplified by introducing rescaled rates, which we define as the ratio, for each species, of entry and exit rates:
\begin{equation}
  \label{eq:rescaledrates}
  \tilde\alpha_{A} = \frac{\alpha_A}{\beta_A}
  \qquad\mbox{and}\qquad
  \tilde\alpha_{B} = \frac{\alpha_B}{\beta_B}
  \ .
\end{equation}

Phases can now be delimited by the transition lines of the phase diagram, as each condition in Eq.~(\ref{eq:tasep:conditions}) excludes a certain phase for a particular zone. For the purpose of illustration, consider the standard TASEP, as represented in Fig.~\ref{fig:tasep:phases}, and focus on identifying the HD phase.
We proceed in two steps. First, according to condition~(\ref{eq:tasep:conditions}-LD-i) the LD phase cannot be present if $\alpha>\beta$, and therefore the region below the line $\alpha=\beta$ can only pertain to an HD or an MC phase. Second, from condition~(\ref{eq:tasep:conditions}-HD-ii) the HD phase cannot occur when $\beta>1/2$, and therefore below the line $\beta=1/2$ we must be dealing with either an $LD$ or a $MC$ phase. Combining these conditions thus identifies the zone in the $(\alpha,\beta)$ plane which corresponds to the HD phase. This way of constructing the phase diagram is graphically represented in Fig.~\ref{fig:TASEP:prevailing}(b).

\begin{figure*}
  \subfloat[$\beta_B<\beta_A<1/2$]{ \label{fig:2pop:sketches:i}
    \includegraphics[height=0.3\textwidth]{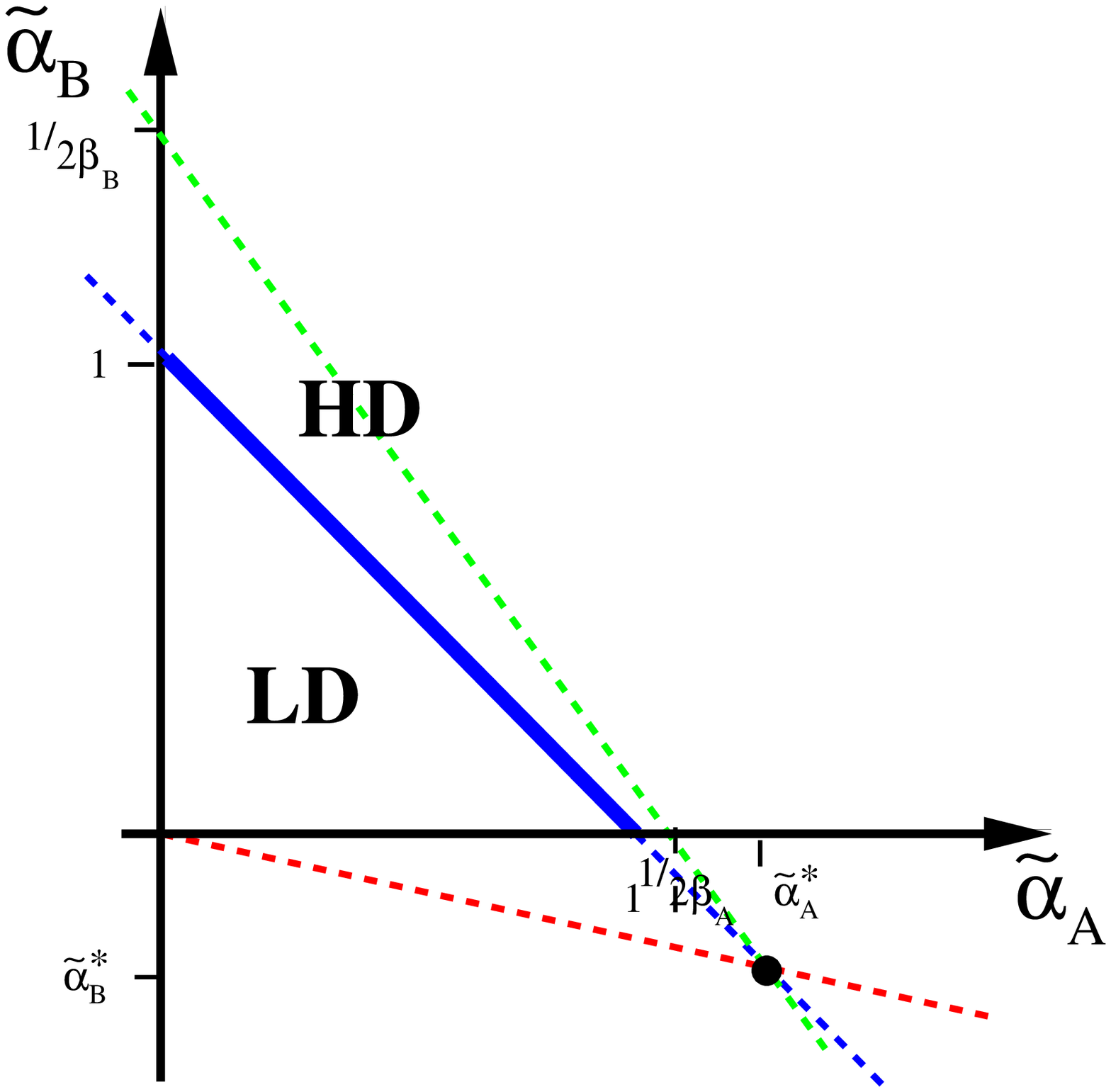}}
  \hfill
  \subfloat[$\beta_B<1/2<\beta_A$]{ \label{fig:2pop:sketches:ii}
    \includegraphics[height=0.3\textwidth]{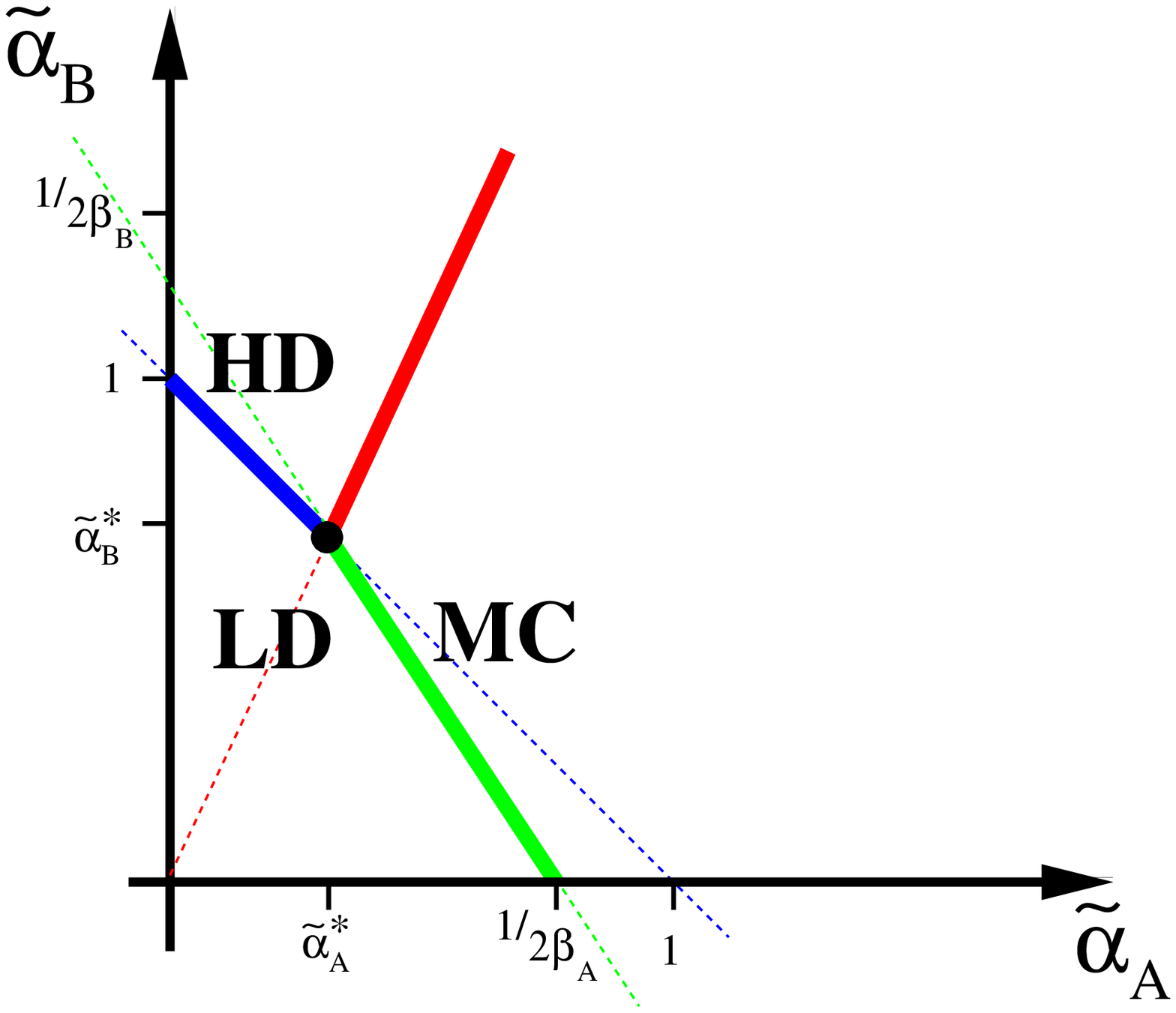}} 
  \hfill
  \subfloat[$1/2<\beta_B<\beta_A$]{ \label{fig:2pop:sketches:iii}
  \includegraphics[height=0.3\textwidth]{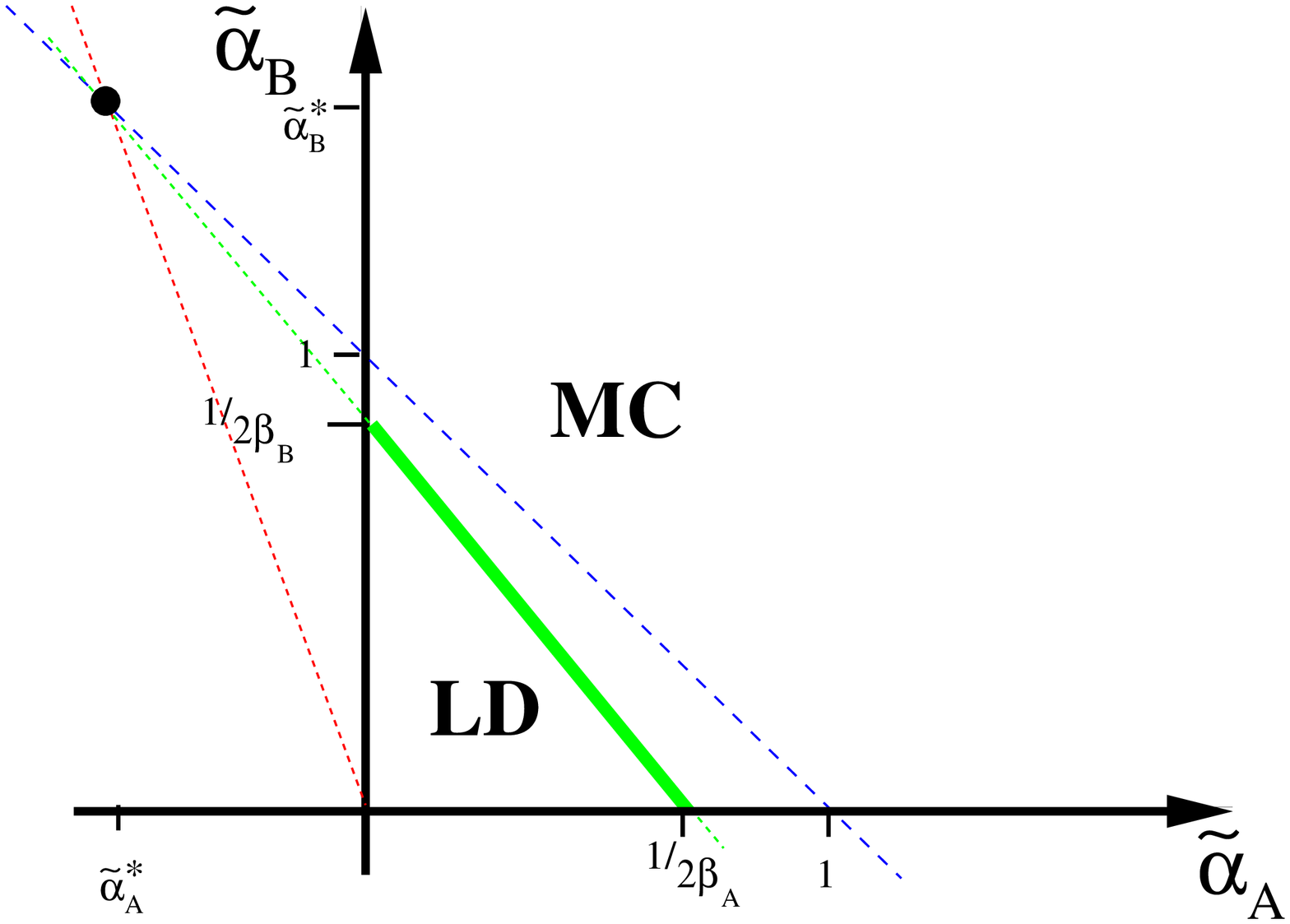}}
  \caption{Mean-field diagram representing two-population TASEP in 3 different scenarios depending the rescaled parameters $\tilde\alpha_A$ and $\tilde\alpha_B$ when $\beta_B<\beta_A$. Scenarios for the phase boundaries are determined by first positioning the critical point (seen in Eq.~( \ref{eq:triplepoint})). We therefore have three cases, which lead to the following phases:   (a)  LD and HD arise where $\beta_B<\beta_A<1/2$,  (b)  LD, MC and HD arise where $\beta_B<1/2<\beta_A$ and finally (c)  LD and MC arise where $1/2<\beta_B<\beta_A$.
}
  \label{fig:pd2pop:sketches}
\end{figure*}

For the full model we can proceed similarly in the ($\alpha_{eff}$,$\beta_{eff}$) plane, as we know that the transitions between phases fall onto (part of) the following relations (colours refer to Fig.~\ref{fig:pd2pop:sketches}):
  \begin{enumerate}
    \renewcommand{\labelenumi}{(\roman{enumi})}
  \item LD-HD: $\alpha_{eff}=\beta_{eff}$ (blue line), which in terms of the rescaled rates is given by
    \begin{equation}
      \label{eq:2pop:condition:HD-LD}
      \tilde\alpha_B=1-\tilde\alpha_A.
    \end{equation}
  \item LD-MC: $\alpha_{eff}=\frac{1}{2}$ (green line) or, equivalently,
    \begin{equation}
      \label{eq:2pop:condition:MC-LD}
      \tilde\alpha_B=\frac{1}{2\beta_B}-\frac{\beta_A}{\beta_B}\tilde\alpha_A.
    \end{equation}
  \item HD-MC: $\beta_{eff}=\frac{1}{2}$ (red line), or equivalently
    \begin{equation}
      \label{eq:2pop:condition:MC-HD}
      \tilde\alpha_B=-\frac{1-2\beta_A}{1-2\beta_B}\tilde\alpha_A.
    \end{equation}
\end{enumerate}
  As a direct conclusion, all phase boundaries are straight lines in the $(\tilde\alpha_A,\tilde\alpha_B)$ plane.
  Fig.~\ref{fig:pd2pop:sketches} shows how by combining these conditions we can assign a zone to each phase in the $(\tilde\alpha_A,\tilde\alpha_B)$ plane.
  First of all, the LD-HD line is fixed in this representation, and goes through $(0,1)$ and $(1,0)$. Next, the LD-MC boundary intersects the axes at $(1/(2\beta_A),0)$ and  $(0,1/(2\beta_B))$. Finally, the point where all these lines cross is located at
\begin{equation}\label{eq:triplepoint}
  (\tilde\alpha_A^*,\tilde\alpha_B^*) = \Big(\frac{1/2-\beta_B}{\beta_A-\beta_B},\frac{\beta_A-1/2}{\beta_A-\beta_B}\Big)
  \ .
\end{equation}

In the following we assume the B particles to be the ones with a slow exit rate, i.e. we take $\beta_B < \beta_A$ (without restricting generality, as the opposite case would be covered by exchanging particle species).  Three different scenarios can now easily be identified, according to how the exit rates compare to the threshold of $1/2$. This yields three cases:
\begin{itemize}
\item (a) $\beta_B<\beta_A<1/2$,
\item (b) $\beta_B<1/2<\beta_A$, and
\item (c) $1/2<\beta_B<\beta_A$. 
\end{itemize}
These cases differ in the relative positions at which the LD-HD and LD-MC separation lines intersect the $\tilde\alpha_B$ axis, while the LD-HD line remains fixed. The construction of the phase diagram can be visualised most clearly if one first admits negative values for the rates $\tilde\alpha_A$ and  $\tilde\alpha_A$ before restricting our interpretation to the physically relevant area. With this in mind, the triple point  $(\alpha_A^*,\alpha_B^*)$ may be localised in various quandrants of the plane, and it is this which distinguishes the three scenarios. The three cases are illustrated in  which correspond to subfigures \ref{fig:pd2pop:sketches} (a-c) .

From this construction it follows that the LD  phase is assigned to the area which is both below the blue and the green lines, and this can be achieved in all three scenarios.
HD is delimited by the blue and the red lines. In scenario (a) and (b) it corresponds to the zone above both the blue and the red line. In scenario (c) however it corresponds to the one above the blue and below the red line: these conditions cannot be met in the physical domain (positive rates), such that there is no HD phase whenever $1/2<\beta_B<\beta_A$.
Finally, the area corresponding to MC is delimited by the green and the red lines.
 In scenarios (a) and (b) this is the zone above the green line and below the red line. In case (a), however, this zone is not part of the physical region, and so there is no MC phase whenever $\beta_B<\beta_A<1/2$. In scenario (c), the area corresponding to MC is above the green and the red lines.
These results are compared to numerical simulations in Fig. \ref{fig:pd2pop}, as discussed in the next subsection.

\subsection{Interpretation}

Figs~\ref{fig:2pop:sketches:i}, \ref{fig:2pop:sketches:ii} and \ref{fig:2pop:sketches:iii} illustrate the different scenarios of the mean-field diagram representing two-population TASEP. The arguments that follow are generic, no correspondence with data is sought at this stage. The scenarios can be distinguished based on the location of the critical point (Eq.~(\ref{eq:triplepoint})) in the plane of rescaled in-rates $(\tilde\alpha_A,\tilde\beta_B$): this essentially fixes all the phase boundaries meeting here,  all of which are straight lines (see Eqs.~(\ref{eq:2pop:condition:HD-LD}, \ref{eq:2pop:condition:MC-LD} and \ref{eq:2pop:condition:MC-HD}).
The LD phase is present in any scenario, as expected, since this is the 'default' phase which can always be reached by sufficiently lowering all input rates.
All three phases are observable if the critical point falls into the physically accessible parameter domain (i.e. into the first quadrant, panel (b)). If it falls into one of the adjacent quadrants, however, only two phases are observable (LD and HD in panel (a), or LD and (MD) in panel (c)).
The colour code for the phase boundaries is that of Fig. \ref{fig:tasep:conditions}; the solid part of the lines indicate the actual phase boundaries. 
Focusing on Fig. \ref{fig:2pop:sketches:i} first shows that an HD phase is possible, and it is in fact the only other phase in the physical region (positive rates), if the fast exit rate is sufficiently small ($\beta_A<1/2$): this is necessary, and sufficient, to limit the out-flow and to provoke a high density throughout the segment. In this case the HD phase will be found in the system if the inflow of particles is sufficiently large, and interestingly this criterion is not given directly by the in-rates, but rather in terms the ratio of in-rate to out-rate for each species (see Eq.~(\ref{eq:rescaledrates})).

Conversely, Fig.  \ref{fig:2pop:sketches:iii} shows that an MC phase is present in the phase diagram if the slower exit rate is sufficiently large ($\beta_B>1/2$), as then the exit current is sufficient to keep particles from building up in an HD zone.

Finally, there is an intermediate regime:  if the slow exit rate is sufficiently small ($\beta_B <1/2)$ while the fast exit rate is sufficiently large ($1/2<\beta_A$), both an HD and an MC phase are possible, see  Fig.  \ref{fig:2pop:sketches:ii}. Indeed, increasing the proportion of fast exiting particles (i.e. increasing $\alpha_A$) will push the system into an MC phase, whereas having more slow particles (i.e. increasing $\alpha_B$) will favour an HD phase.

A slightly contrasting statement is to be made in terms of which of the parameters are decisive. Indeed, within each scenario (a, b or c), the phase can be identified based solely on the reduced entry rates $\tilde\alpha_{A}$ and $\tilde\alpha_{B}$, as illustrated in the phase diagrams. However, it is worth noting that {\it  all} rates ($\alpha_{A}$ and $\alpha_{A}$ as well as $\beta_{A}$ and $\beta_{B}$) are required explicitly in order to determine which regime the system finds itself in.

\subsection{Numerical validation}

Stochastic numerical simulations were performed with the Gillespie algorithm \cite{Gillespie:77}. This consists in picking, iterating over time,  one of all possible changes in the system which may occur, with the appropriate statistical weights. Each such ``move''or ``rection'' is attributed a timescale, which is drawn from the corresponding waiting time distribution, therfore implementing the time evolution of the system. A lattice of length $L = 500$ sites was used and, unless stated otherwise, measurements were cumulated over $10^8$ Gillespie iterations, after having discarded a transient of $4\times10^7$ iterations.


\begin{figure*}
  \subfloat[$\beta_A = 0.4$ and $\beta_B = 0.2$]{\includegraphics[width=0.33\textwidth]{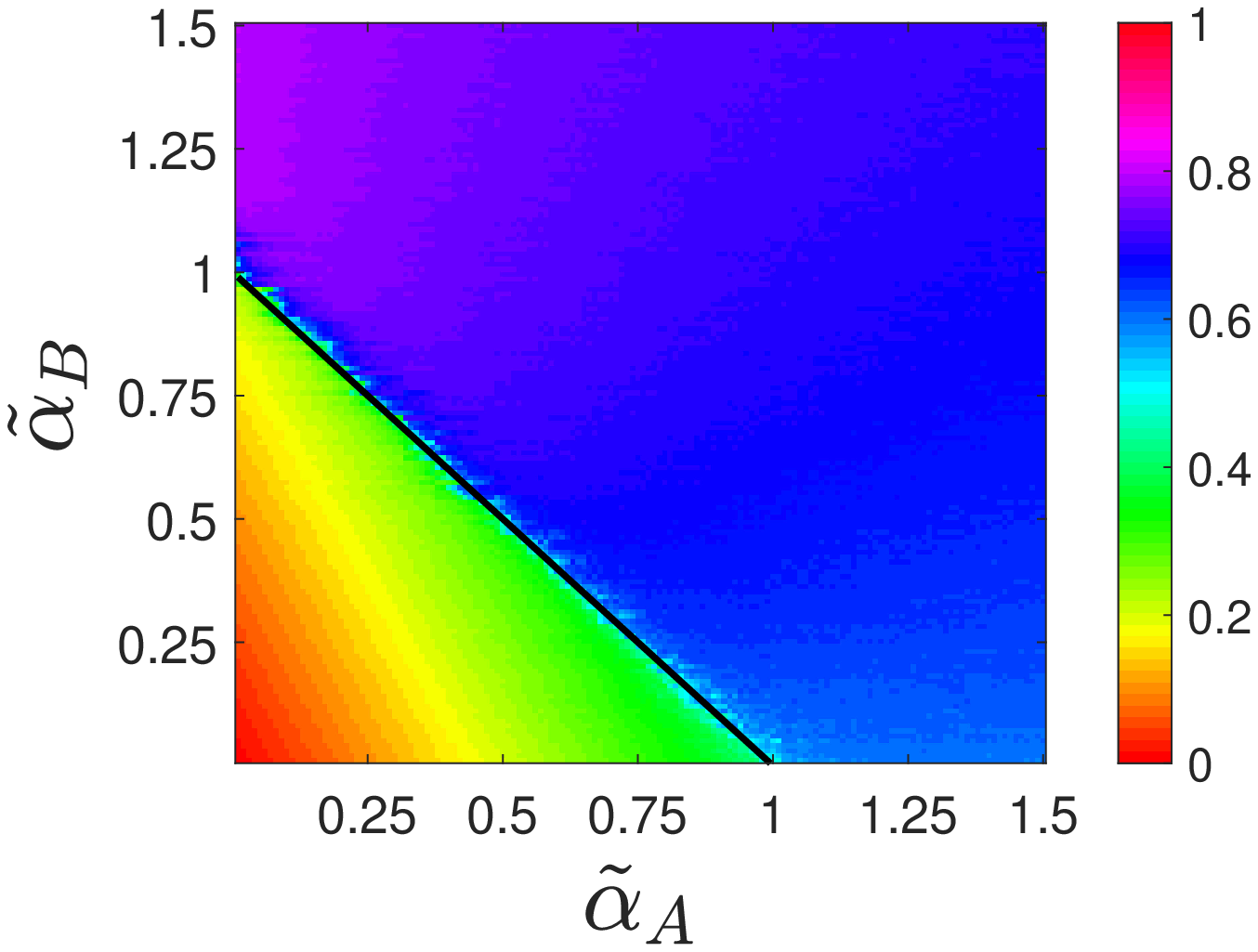}}
  \hfill
  \subfloat[$\beta_A = 0.625$ and $\beta_B = 0.333$]{\includegraphics[width=0.33\textwidth]{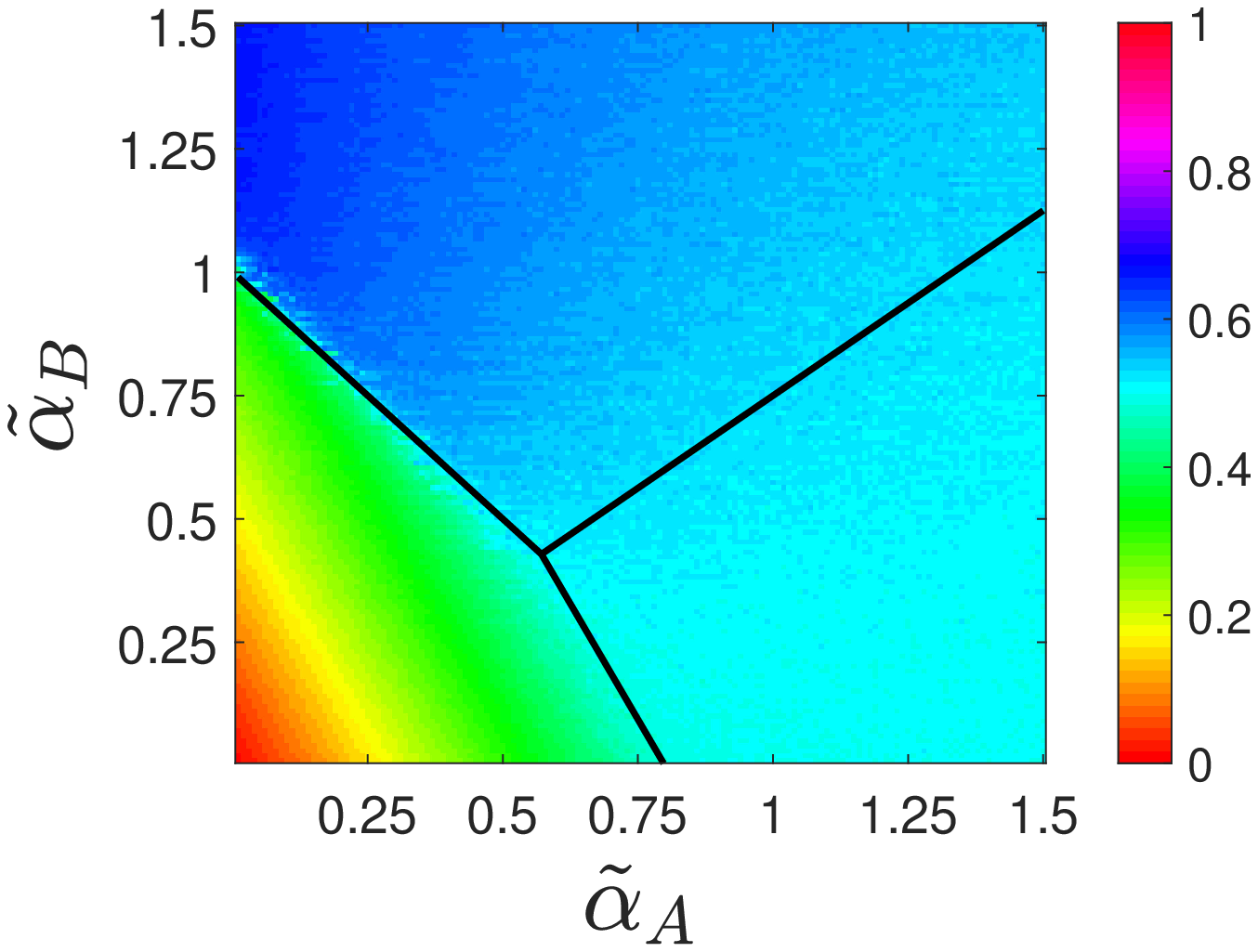}}  
  \hfill
  \subfloat[$\beta_A = 0.666$ and $\beta_B = 0.625$]{\includegraphics[width=0.33\textwidth]{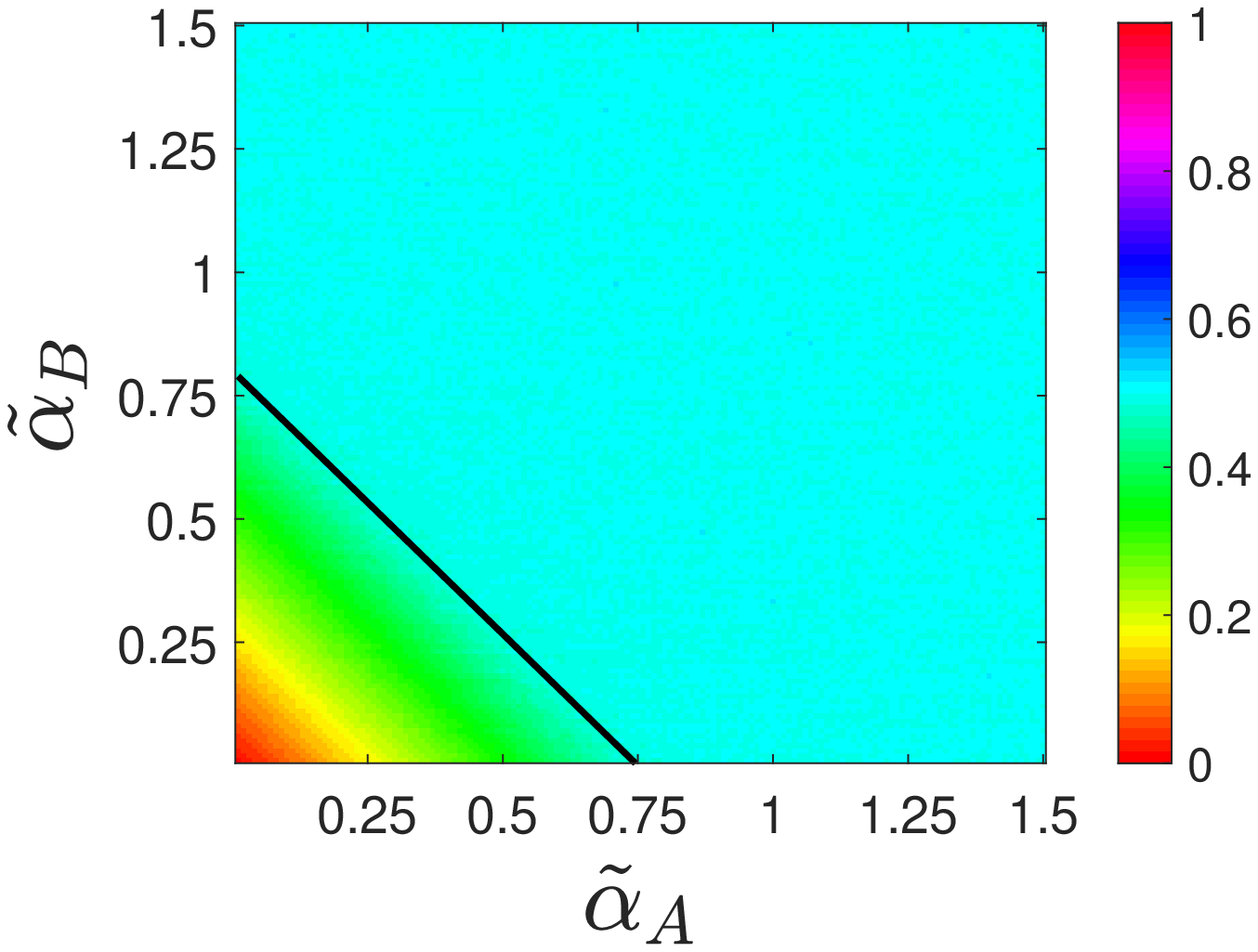}}
  \caption{
Scenarios for the phase diagram from numerical data, plotted in the same rescaled parameter plane $(\tilde\alpha_A,\tilde\alpha_B)$ as used in the previous figure. The colour gradient plot represents the average density throughout the lattice, calculated on a grid $(\tilde\alpha_A,\tilde\alpha_B) \in [0.01,1.5] \times [0.01,1.5]$ with a step of $0.01$. The black lines in panels (a), (b) and (c) correspond to the analytical expressions for the phase boundaries, as given by Eqs.~(\ref{eq:2pop:condition:HD-LD}), (\ref{eq:2pop:condition:MC-LD}) and (\ref{eq:2pop:condition:MC-HD}).}
 \label{fig:pd2pop}

\end{figure*}

In order to confront the analytical mean-field characterisation of phases to data from simulations we plot in Fig. \ref{fig:pd2pop} the numerically obtained density (averaged both over time and the segment), from which the phases can be deduced ($\rho=1/2$ for MC, below or above this value for LD and MC, respectively). Analytical expressions for the boundaries between the phases are superposed, showing a very good correspondence. For the chosen set of parameters the theoretical analysis is therefore validated.

\begin{figure}
  \centering
  \includegraphics[width=0.5\textwidth]{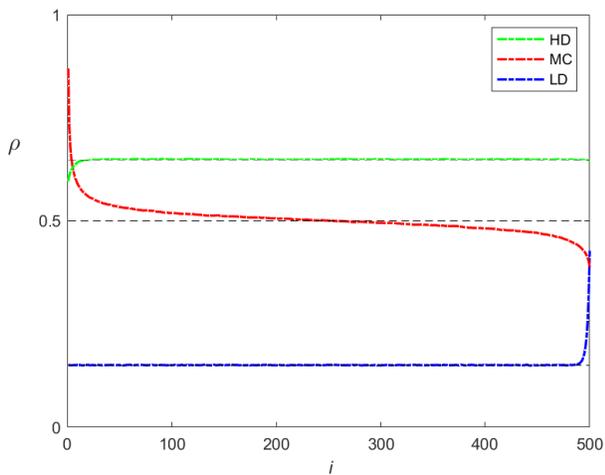} 
   \caption{
    \label{fig:mf:profiles}
    Numerically acquired density profiles (coloured lines), contrasted to the mean-field predictions (black dashed line).    The following phases are shown:
    LD ($\alpha_A=0.1$, $\beta_A=0.4$; $\alpha_B=0.05$, $\beta_B=0.2$) (blue line);
    MC ($\alpha_A=1$, $\beta_A=2./3$; $\alpha=0.9375$, $\beta_B=0.625$) (red line);
    HD ($\alpha_A=0.0625$, $\beta_A=0.625$; $\alpha_B=0.5$, $\beta_B=1/3$) (green line).
    We obtain good agreement for these choices of parameters, except for the boundary layers which are expected from the regular single-species TASEP. 
    For other parameter choices deviations arise, which are discussed below and analysed in the following section.
     }
  \end{figure}

For a finer comparison we also contrast the density profiles along the segment to the analytical mean-field prediction.  Figure \ref{fig:mf:profiles} shows numerical data for three sets of parameters, which correspond to examples for LD, MC and HD phases. We compare to mean-field where the  predictions are based on the mapping onto a single-species TASEP via the appropriate effective rates (see Eqs.~(\ref{eq:effective:alpha}) and (\ref{eq:effective:beta})). Again, we obtain good correspondence in that the value of the bulk density is well predicted. Deviations occur only at the boundaries, as is indeed expected: such boundary layers are known to arise for the regular TASEP model, and therefore do not constitute a new feature of the 2-species model. Thus the LD  phase presents a boundary layer close to the exit point, the HD  phase close to the entry point, and the MC phase at both ends, while preserving a slight overall slope in the density profile, which is known to vanish in the limit of infinite system size \cite{krapivsky}.

\subsection{Mean-field discrepancies}

\begin{figure}
  \subfloat[]{\includegraphics[width=0.5\textwidth]{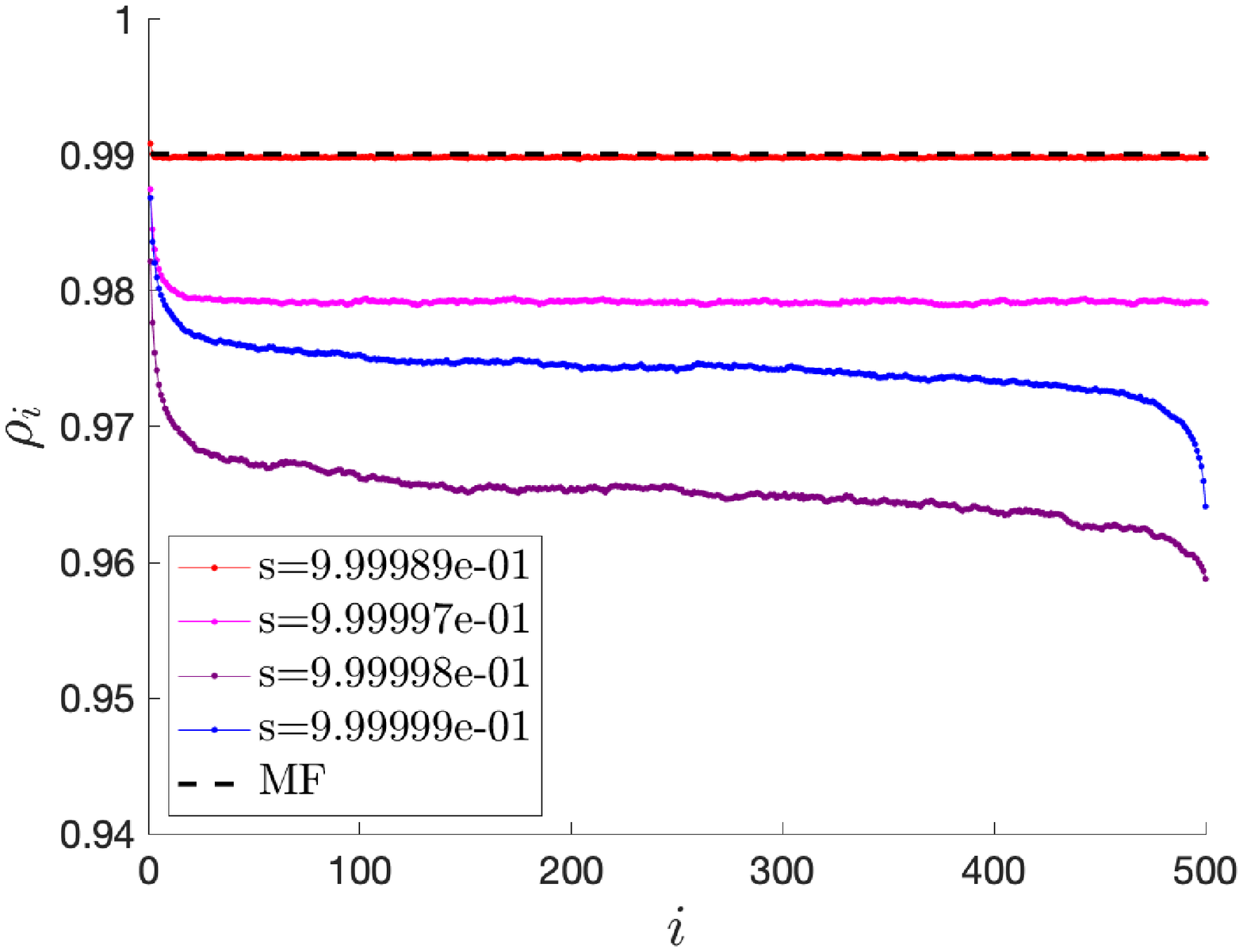}}
  \hfill
\subfloat[]{\includegraphics[width=0.5\textwidth]{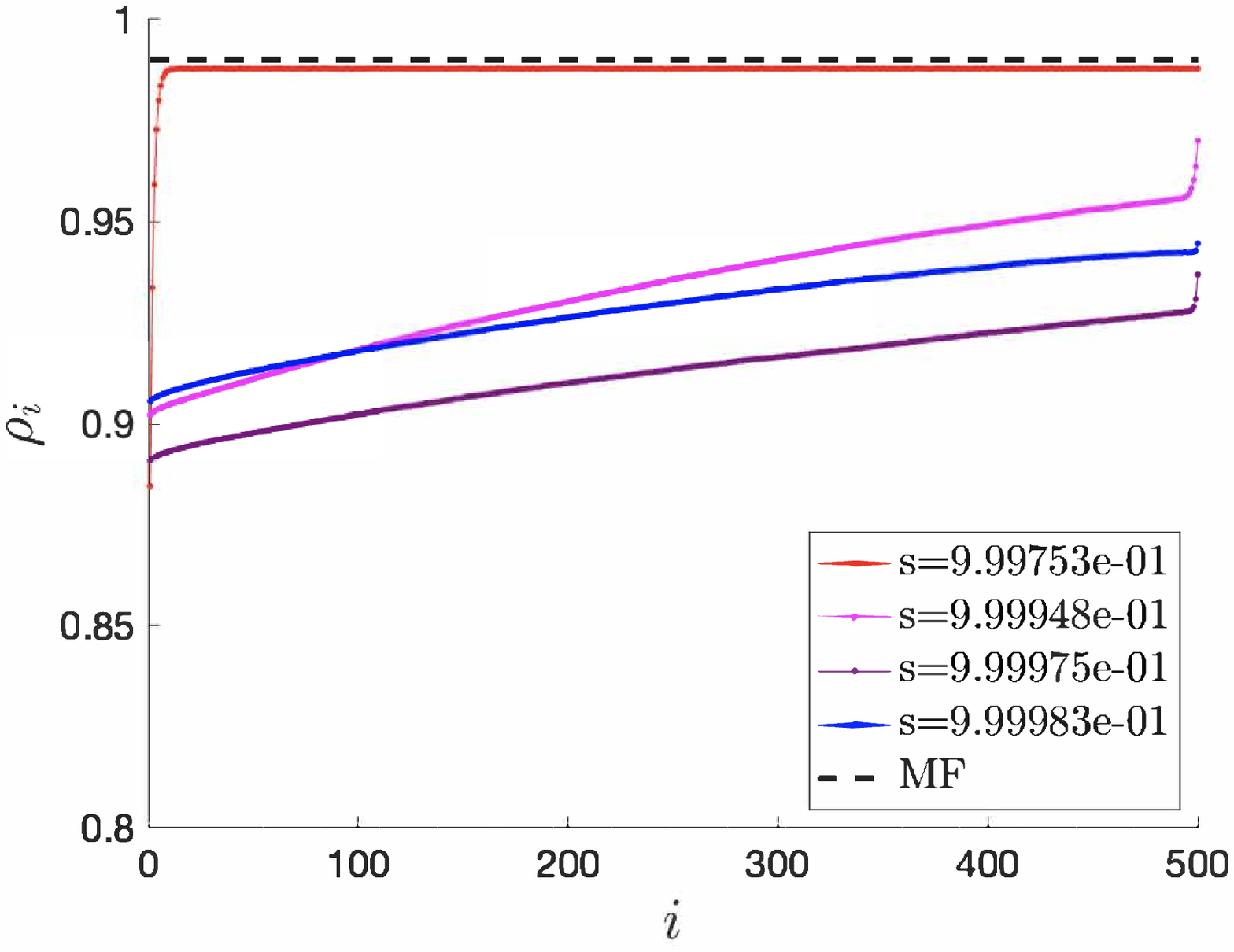}}  
  \caption{
    \label{fig:intermittency:motivation}
    HD density profiles from numerical simulation, illustrating the breakdown of mean-field predictions in the case where the exit rates $\beta_A$ and $\beta_B$ differ greatly. For all simulations   parameters were chosen as to maintain identical effective rates, just as the proportion of B particles ($\chi_B$). The plots represent several choices for the slowness parameter, $s =1- \beta_B/\beta_A$. The single-species prediction is thus seen to break down as the species B becomes increasingly slow to leave.
    Panel (a): $\alpha_{eff}=1$, $\beta_{eff}=9.991\times 10^{-3}$, $\chi_B=1\times 10^{-4}$.
    Panel (b): $\alpha_{eff}=1.001\times 10^{-1}$, $\beta_{eff}=9.991\times 10^{-3}$, $\chi_B=1\times 10^{-3}$. In these simulations $10^5$ fast particles lattice (and therefore, on average, 100 slow particles) entered the lattice. Average were calculated over a time of $1 \times 10^7$, after a transient of $10^6$ initiation events (corresponding on average to a time of $7 \times 10^4$), which has been discarded.
  }
\end{figure}

However, as illustrated in Fig. \ref{fig:intermittency:motivation}, this agreement is not fully general: density profiles can differ significantly from the prediction of the effective single-species TASEP model.
The 2-species model can therefore display new features, which are not captured by the  mean-field description in terms of an effective single-species TASEP which we have elaborated so far, at least for certain choices of parameters. In order to qualify these differences we focus again on density profiles, rather than on the entire phase diagram, which will also help to establish a strategy for improving the theoretical approach.

In principle there are 4 independent parameters to the model (two rates for each species), and therefore, the parameter space to be explored is vast.
Varying any of these rates may affect current and density, and may also push the system across a phase boundary. However, the effective one-species model suggests that it is exclusively the two effective rates $\alpha_{eff}$ and $\beta_{eff}$ which determine the behaviour, or at least so long as the mean-field analysis remains valid. We therefore choose, in a first instance, to vary all independent rates, $\alpha_A$ and $\beta_A$ as well as $\alpha_B$ and $\beta_B$, {\it jointly}, such as to preserve the total effective rates. In this way we can compare results from numerical simulations to analytical mean-field predictions without modifying the position in the mean-field phase diagram, i.e. no mean-field phase transition can be triggered by such a change.

We thus need to pick two additional parameters, in adition to $\alpha_{eff}$ and $\beta_{eff}$, to define our system. Before choosing how to do this, consider two limiting cases of our system, one where $\alpha_B=0$ (i.e. the case where we recover the single-species model, since B-particles are absent), and another one where $\beta_B\to\beta_A$ (an equivalent scenario, since both particle species behave identically). In both limits the single-species TASEP model must hold.
This suggests choosing the two remaining parameters in such a way that they characterise (i) the fraction of (slow) B particles in the system and (ii) the 'slowness' of B particles as compared to A particles.

A natural choice for the first parameter thus is the fraction of B particles, $\chi_B \in [0,1]$. Recall that fixing $\chi_B$ directly implies the individual input rates as (see Eq.~(\ref{eq:chiAB:def})) as
\begin{equation}
  \alpha_{A} = (1-\chi_B) \, \alpha_{eff}
  \qquad\mbox{and}\qquad
  \alpha_B = \chi_B \, \alpha_{eff}
  \ .
\end{equation}
Varying $\chi_B \in [0,1]$ maps out this degree of freedom at a fixed total effective rate $\alpha_{eff}$, as desired.

We now pick a second parameter, say $s \in [0,1]$, to play a similar role for the exit rates. Specifically, we require $s$ to make B particles slower, by setting
\begin{equation}
  \beta_B = (1-s) \beta_A
  \ ,
\end{equation}
while requiring that the effective exit rate $\beta_{eff}$ remain unaffected. According to Eq. (\ref{eq:betaeffinverse}) this implies
\begin{equation*}
  \frac{1}{\beta_{eff}} = \frac{\chi_A}{\beta_A} + \frac{\chi_B}{(1-s) \, \beta_A}
\end{equation*}
which can be solved to yield
\begin{eqnarray}
  \beta_A &=& \beta_{eff} \, \left[ \frac{1-\chi_B \,s}{1-s} \right]
  \\
  \beta_B &=& \beta_{eff} \,  \left[ 1 - (1-\chi_B) \, s \right]      
  \ .
\end{eqnarray}
In essence, we can thus use the parameters $\chi_B$ and $s$ to vary the abundance of $B$ particles and their 'slowness' independently, while leaving the effective entry/exit rates unchanged. 

Figure \ref{fig:intermittency:motivation} shows density profiles for two examples for which the mean-field prediction is an HD phase. For all graphs in each panel, the effective in/out rates as well as the particle distribution have been kept constant: the only parameter which is varied is $s$, which regulates the slowness of B particles. Deviations from the mean-field prediction (black dashed line) become increasingly significant as the slowness $s$ increases: since all other parameters have been maintained constant, we can conclude that the mean-field theory fails as B particles become too slow to exit.

Deviations concern not only the average density value, but also the shape of the density profile can deviate significantly from what is expected from an effective single-species description. In Fig. \ref{fig:intermittency:motivation} (a) the shape of the density profile for the two largest values of $s$ resembles a density profile in the maximal current (MC) phase, despite the lattice being in HD phase (average density on the lattice is above 0.5).  In panel (b) the systematic positive slope in the profile makes it qualitatively different from a single-species profile.

In essence, these examples show that the effective single-species model is no longer appropriate as one of the particle species becomes significantly slower to leave than its counterpart. The intuition at this point is that those particles provoke temporary blockages, leading to 'intermittent' flow with entirely new characteristics. We pursue this thought further in the following section, and show how intermittency may be used to construct an improvement to the mean-field predictions.

\section{Intermittency}

In the previous section numerical evidence has exposed the fact that an effective single-species model no longer does justice to the traffic in our two-species model when one particle species becomes very slow to leave. This suggests a considerable alteration of the traffic, the nature of which becomes clear by considering the limiting case where B particles become extremely slow to leave (slowness $s \simeq 1$). One then anticipates having a blocked system whenever a B particle reaches the last site of the lattice. As soon as the B particle exits the lattice, a stretch of A particles will evacuate until the next B particle reaches the exit. Rather than a process of continuous flow, we are therefore looking at periods of flow of particles, interrupted by periods during which the exit is fully blocked. We will refer to this as {\it intermittency} in the following, and show how mean-field arguments can be amended to account for this phenomenon.

\begin{figure*}
    \includegraphics[width=\textwidth]{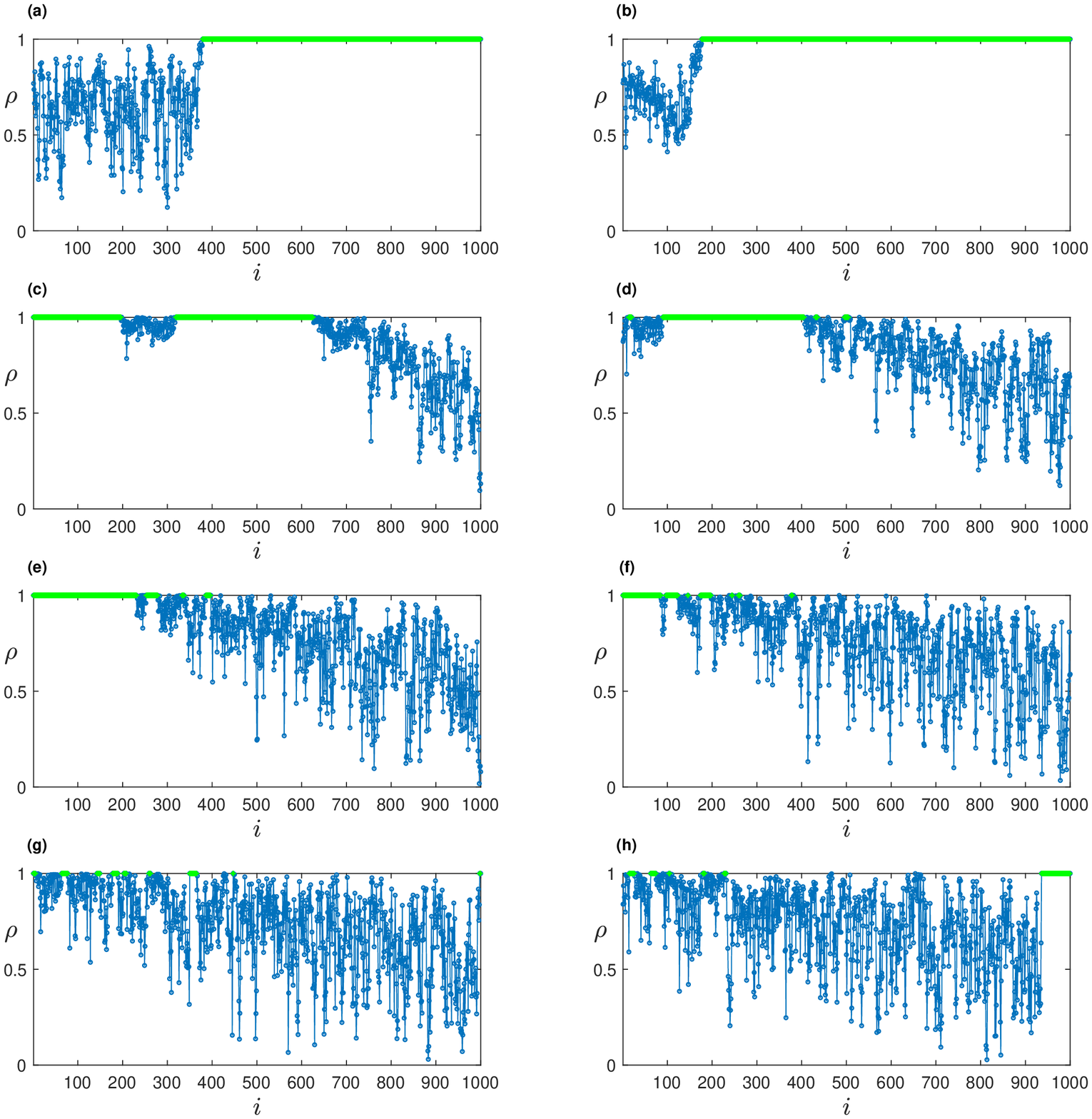}
  \caption{
    \label{fig:intermittency:timeseries}
   Time evolution of the density profiles for the $MC\to HD^{*}$ scenario, showing a temporary blockage of the exit site and its evolution for a lattice of $N=1000$ sites. Panels (a) to (h) represent the succession of instantaneous-like density profiles along the same simulation. Parameters are $\alpha_A = 1$; $\alpha_B =0.007$; $\beta_A = 1$; $\beta_B = 0.003$. Each snapshot has been calculated by averaging over 2,600 Gillespie iterations (each iteration corresponding to one reaction occurring in the system, i.e., the movement of one particle on the lattice). The number of Gillespie iterations separating successive snapshots is equal to 20,800. The green dots are representing an average density equal to 1 over the 2,600 Gillespie iterations and thereby highlighting the bloackage created by the B particle. Successive snapshots are ordered in time but are not separated by identical time intervals: they have been selected to illustrate the essence of the process with only a few snapshots.}
\end{figure*}

In order to better appreciate the phenomenon we show a series of snapshots of density profiles in Fig.~\ref{fig:intermittency:timeseries}.
Each of these is a quasi-instantaneous density profile, obtained by averaging the occupancy of each lattice site over 2,600 Gillespie iterations. 
The 12 graphs are presented in chronological order (from left to right, top to bottom), thus illustrating the time evolution of the density profile. 
Panel (a) shows a blockage at the exit, with a 'jammed' region (highlighted in green). As time progresses, this 'jammed' region 'travels upstream', i.e., to the left, as shown in panel (b). More precisely, it 'grows' to the left, as particles within the jammed region are of course essentially stuck and therefore static, but further particles join the jam from the left. In panel (c) the blocking particle has finally exited the lattice; therefore the particles from the right boundary of the jammed region can start moving ahead, and eventually leave the lattice. Thus the jammed region decreases in size from its right boundary. The net effect is that the jammed region appears to travel upstream, as particles join at its left boundary and others leave the jammed region at its right boundary. Some more complicated effects can occur, as shown in panel (c), where the jammed region breaks into two parts as it 'travels' upstream. In panels (d)-(f) the jammed region is dissolved as it reaches the left boundary of the lattice, and the bulk density relaxes to the one determined  by the faster, non-blocking A particles. 
Ultimately, we can see another jam forming in panel (h), caused by the arrival of a B particle to the last site of the lattice (jammed region highlighted again  in green).

These figures illustrate that the presence of intermittency in the two-population model is the root cause of discrepancy with respect to the effective single-population model, as we will show now.
\\

The simplest case to think about when developing the argument is when the B particles are both very slow to leave ($s \simeq 1$) and  sparse ($\chi_B\ll1$). In this case there will be a simple TASEP current of $A$ particles almost always, except when one of the rare $B$ particles reaches the exit site. Then a jam is created at the exit, corresponding to a stretch of density 1 in front of the exit. However, as soon as the blockage is resolved, the jam evacuates and the system returns to its original phase. As B particles are very sparse, this is essentially a process involving A particles only, and thus the flow phase is in fact characterised by the underlying 'pure' single-species phase (obtained asymptotically as $\chi_B\to 0$ while maintaining all other parameters).  
For example, in Fig. \ref{fig:intermittency:timeseries} we are dealing what appears to be an HD phase which falls back to an MC phase during the periods of continuous flow: we refer to this as a $MC\to HD^*$ phase.
What we mean by this notation is that the underlying 'pure' system of A particless would be found in an MC phase, as is indeed seen during the periods without blockage. However,  due to the presence of slow-to-leave B particles, the resulting phase is more apparent of an HD phase. The star thus labels those phases which already are the result of intermittent behaviour.

\begin{figure*}
  \subfloat[LD$\to$HD$^*$]{  \label{fig:HD-LD:rho}
    \includegraphics[width=0.45\textwidth]{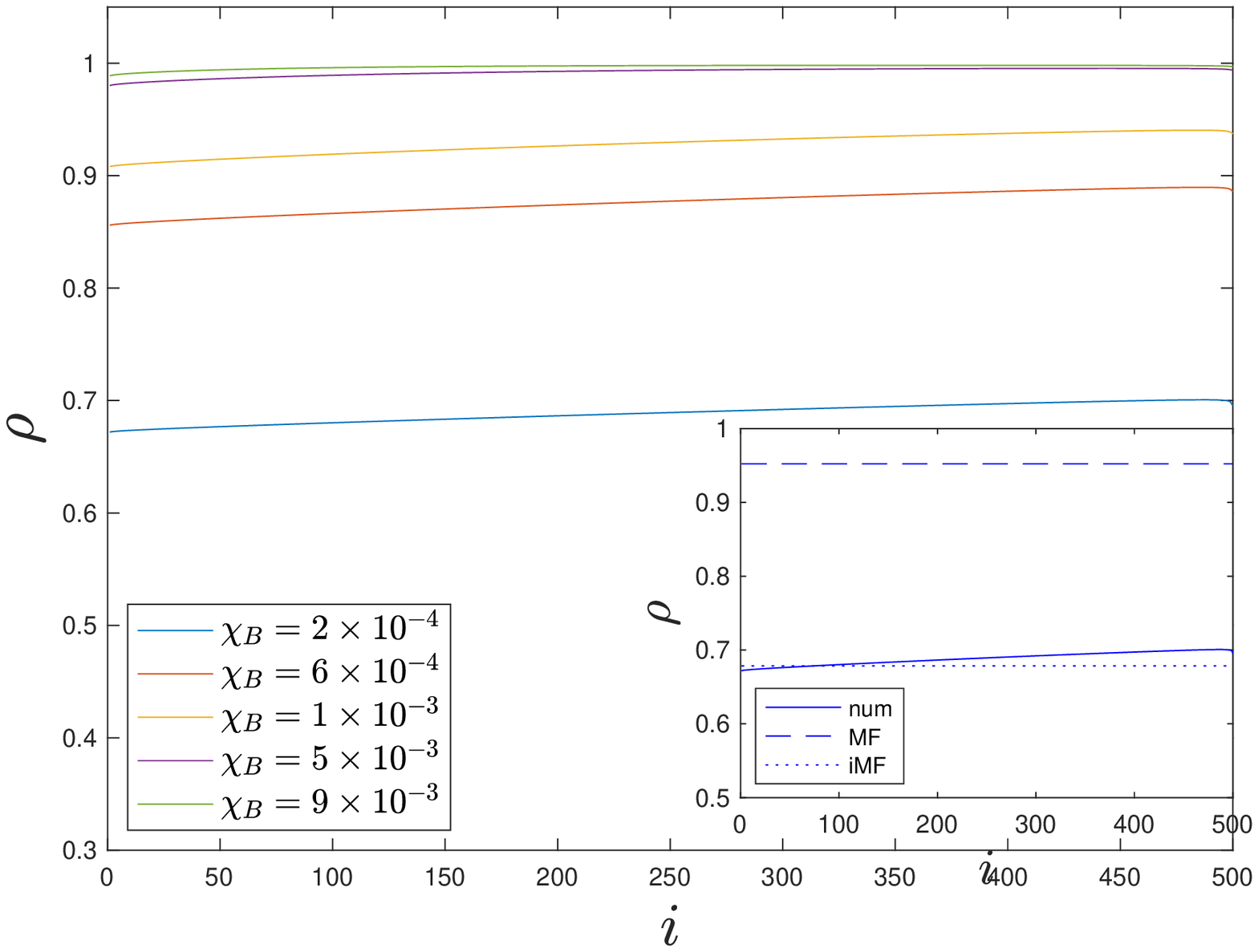}}
  \hfill
  \subfloat[LD$\to$HD$^*$]{  \label{fig:LD-HD:J}
    \includegraphics[width=0.45\textwidth]{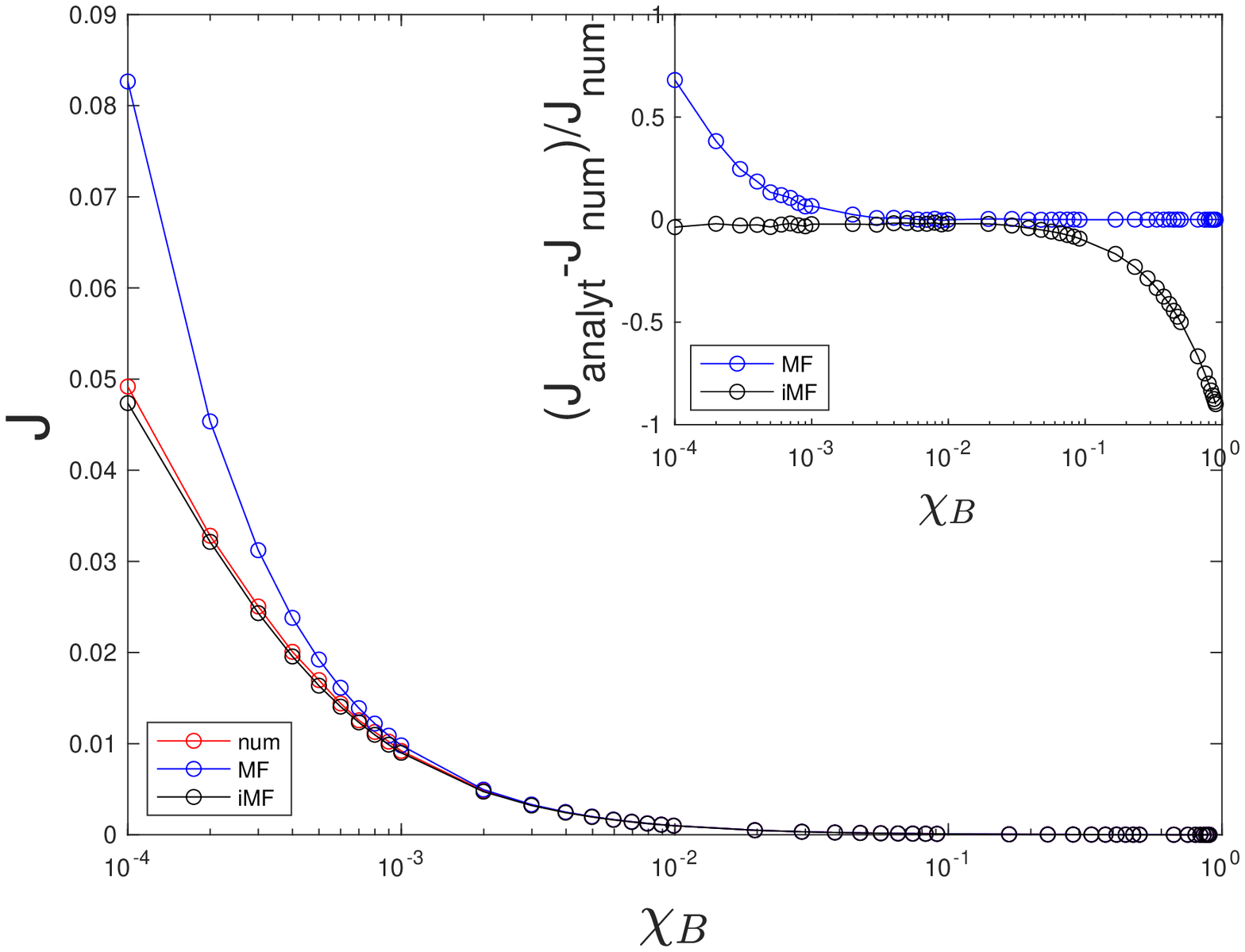}
  }
  \\
  \subfloat[MC$\to$HD$^*$]{  \label{fig:HD-MC:rho}
    \includegraphics[width=0.45\textwidth]{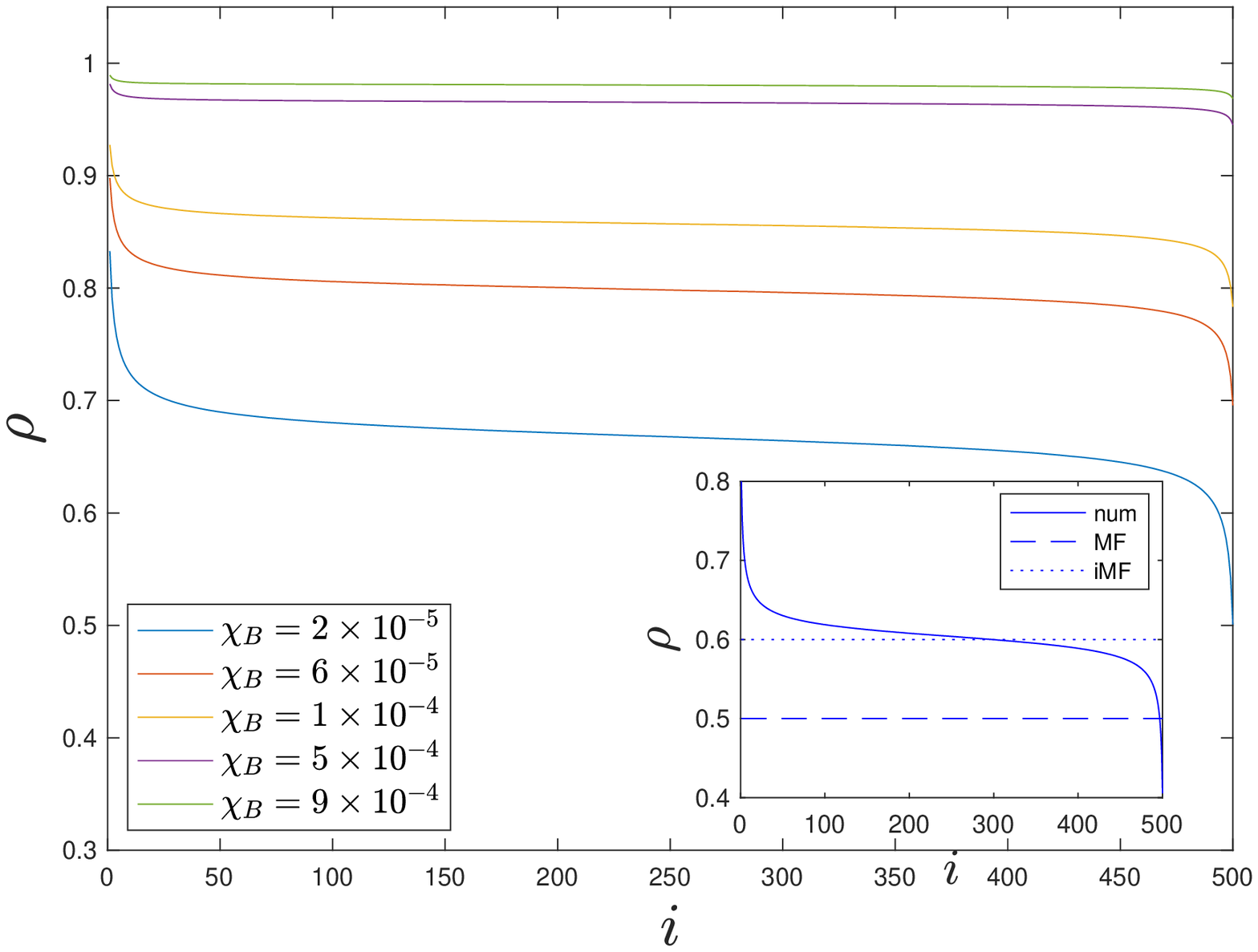}}
  \hfill
  \subfloat[MC$\to$HD$^*$]{  \label{fig:MC-HD:J}
    \includegraphics[width=0.45\textwidth]{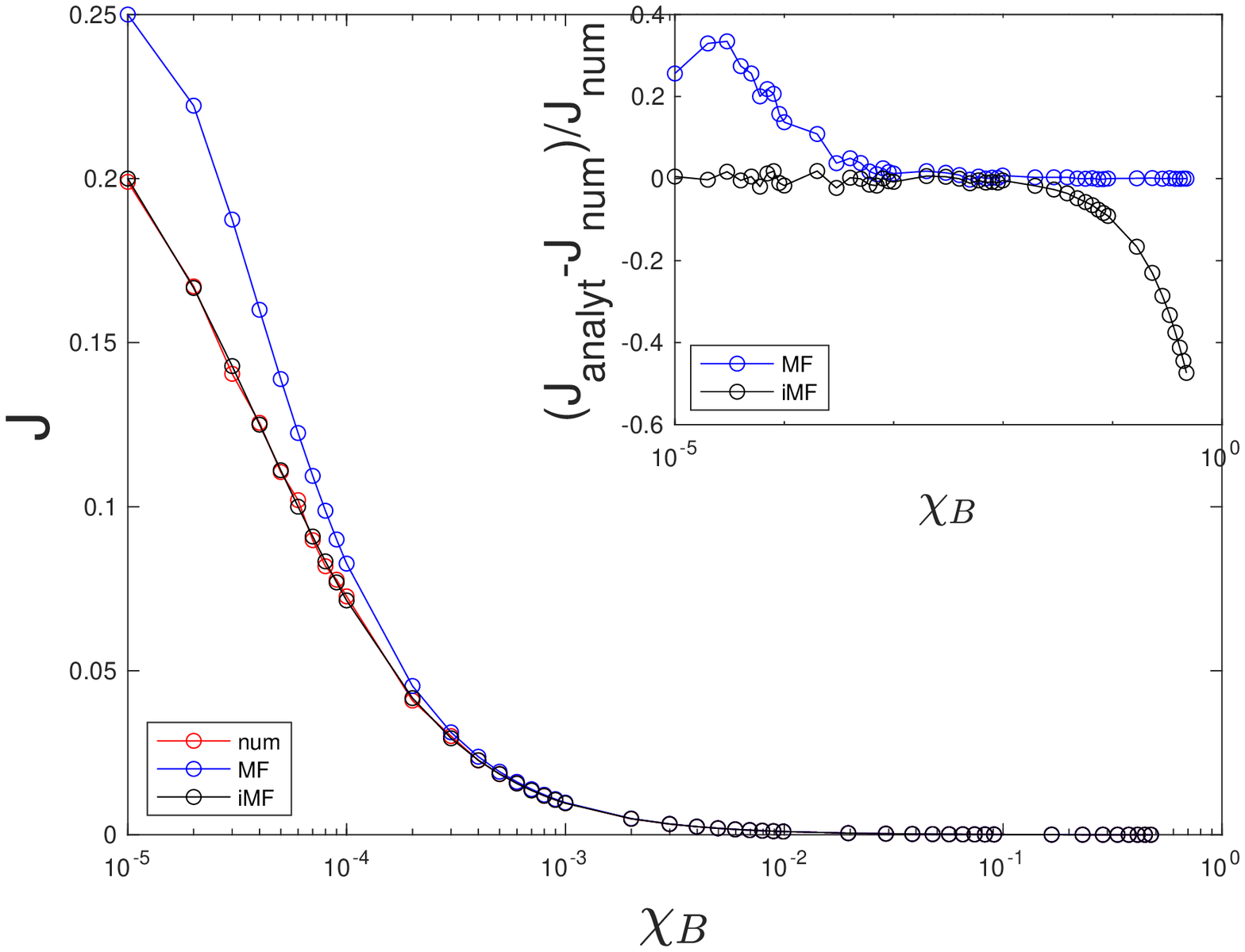}
  }  
  \\
  \subfloat[HD$\to$HD$^*$]{  \label{fig:HD-HD:rho}
    \includegraphics[width=0.45\textwidth]{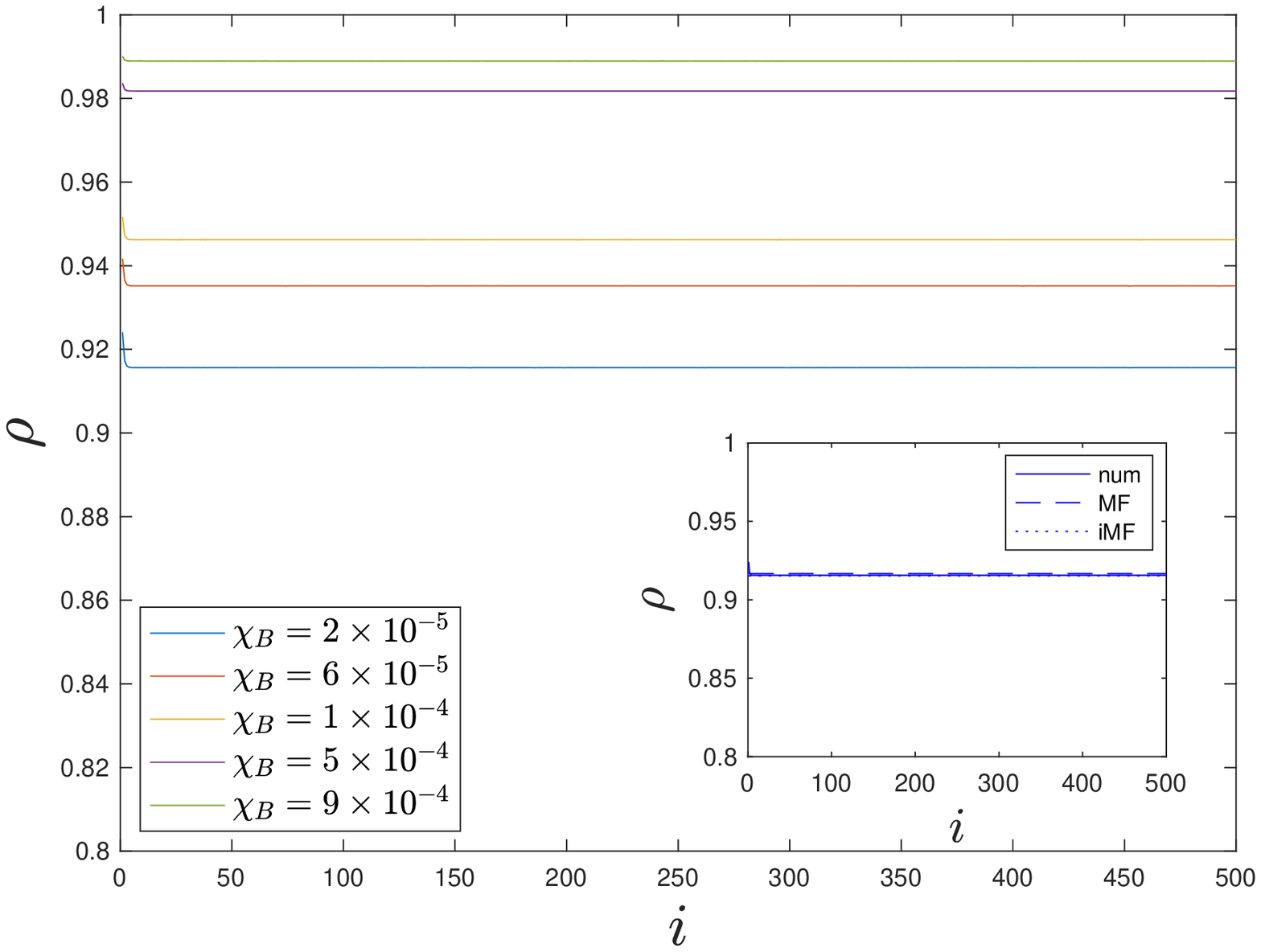}
  }
  \hfill
  \subfloat[HD$\to$HD$^*$]{  \label{fig:HD-HD:J}
    \includegraphics[width=0.45\textwidth]{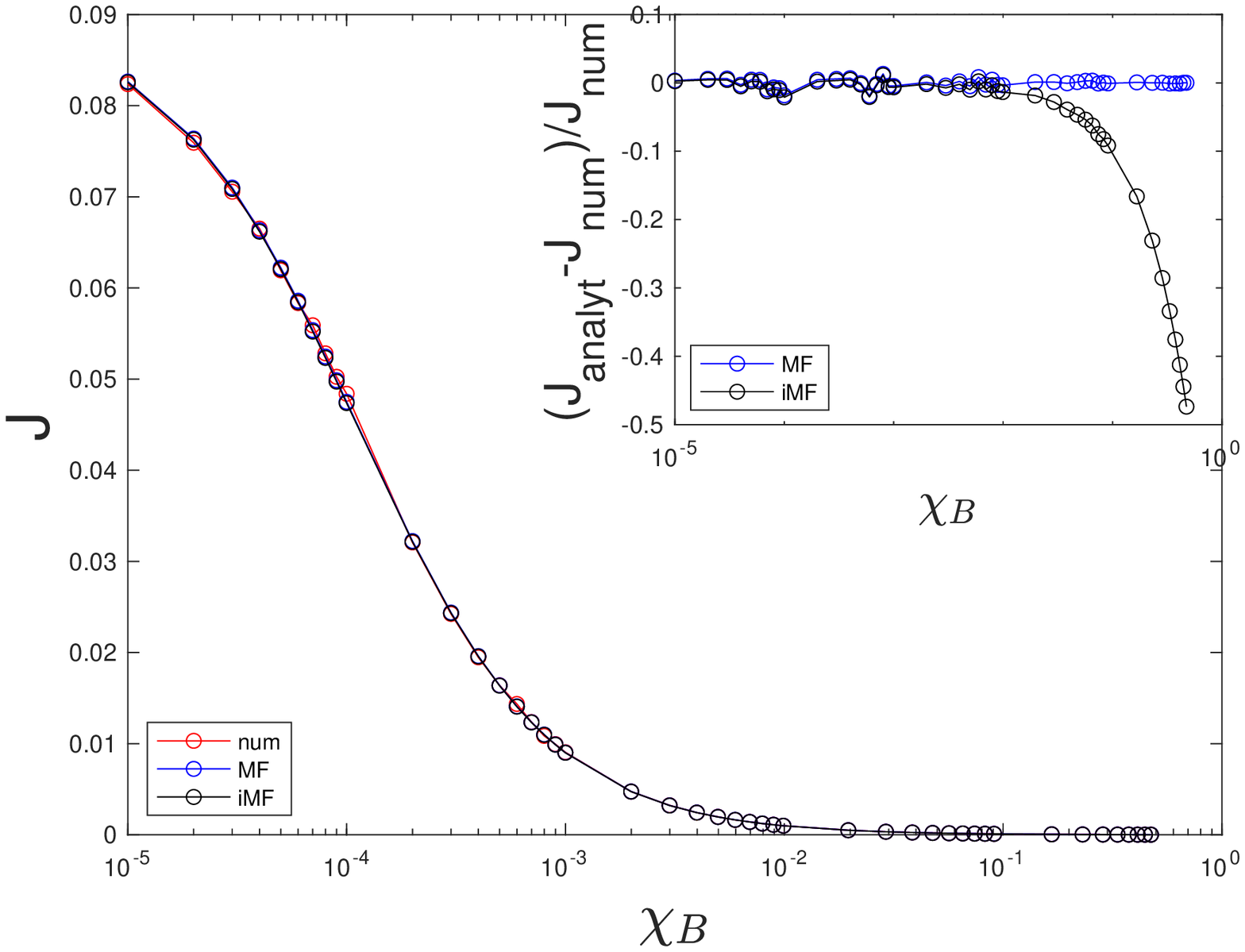}
  }
  \caption{ Effect of intermittency on the density profile $\rho^{(i)}$ (panels a,c,e) and on the current (panels b,d,f), as the proportion $\chi_B$ of slow particles is progressively increased. 
    The parameters used in the sub-figures are as follows:
    (a,b) LD$\to$HD$^*$ phase, with $\alpha_A = 0.1$, $\beta_A = 1$, $\beta_B= 1\times10^{-5}$;
    (c,d) MC$\to$HD$^*$ phase, with $\alpha_A = 1$; , $\beta_A= 1$, $\beta_B = 1\times10^{-5}$;  
    (e,f) HD$\to$HD$^*$ phase, with  $\alpha_A = 1$, $\beta_A = 0.1$, $\beta_B = 1\times10^{-5}$.
    The insets show the results from the numerical simulations (solid line), MF approach (dashed line) and iMF approach (dotted line) for $\chi_B=2\times 10^{-4}$ (panel a); $\chi_B=1\times 10^{-5}$ (panel c); $\chi_B=2\times 10^{-5}$ (panel e).  
    The plots of current $J$ (panels b,d,f) show numerical results (red), MF prediction (blue) and an iMF prediction (black) as a function of $\chi_B$.  The insets show the relative error of the current for both the MF and iMF predictions.
  }\label{fig:intermittency_rho_J}
\end{figure*}

With this picture in mind we now focus on the effect of intermittency in the density profile for 3 different scenarios, corresponding to 3 different choices for the underlying 'pure' single-species phase. These are: LD$\to$HD$^*$ (Fig.~\ref{fig:intermittency_rho_J}a), MC$\to$HD$^*$ (Fig.~\ref{fig:intermittency_rho_J}c) and HD$\to$HD$^*$ (Fig.~\ref{fig:intermittency_rho_J}e). For each of these scenarios we fix the entry and exit rates of the A particles, as well as the exit rate of the B particles, but we vary the proportion $\chi_B$ of particles of type B by changing $\alpha_B$
\footnote{Notice that, consequently, here we are not fixing the effective entry and exit rates for this comparison: although this might have been desirable, it turns out to lead to transitions in the underlying 'pure' phase, e.g. from $LD$ to $ HD$, and therefore complicates the interpretation.}.

Remarkably, in both the LD$\to$HD$^*$ and MC$\to$HD$^*$ cases, the density profiles are qualitatively different from the ones of an effective single species TASEP in an HD phase. In the LD$\to$HD$^*$ scenario, the density profiles exhibit a positive slope from the left to the right boundary of the lattice. In the MC$\to$HD$^*$  scenario the shape of the density profiles resembles the ones of an MC single species TASEP, but with an average density higher than 0.5. The only density profiles that remain qualitatively the same are the ones in the HD$\to$HD$^*$ scenario, although they are quantitatively different from the mean-field predictions. In the next subsection we introduce an extended mean-field approach that accounts some extent for the results obtained in this intermittent regime.

\subsection{Intermittent Mean Field (iMF) Approximation}

We propose to pursue the picture established above to propose an  'intermittent mean-field theory' (iMF) as an extension to the mean-field arguments presented above, designed to account for intermittency in the case where B particles are very slow ($\beta_B \ll \beta_A$) and very sparse ($\chi_B \ll \chi_A$).
We are thus dealing with entire stretches of $A$ particles, say $n_A$ of them on average, separated by isolated $B$ particles. The average number $n_A$ of A particles in each stretch can be then estimated as $\alpha_A/\alpha_B$.
To a first approximation we can therefore think of the current as being the one corresponding to the underlying 'pure' phase (i.e. the phase which corresponds to vanishing $\chi_B$). 
It is interrupted every so often, for the time it takes for the occasional slow $B$ particle to free the exit site. During this period the exit current is zero. As long as these intermittent jams persist, and as an HD-like stretch builds up close to the exit site, we can estimate this average 'blockage' time interval as
\begin{equation}
  \label{eq:tauHD}
 \tau\bl = \frac{1}{\beta_B} \ .
\end{equation}
Clearly, for this to be valid, we must require that transient periods are short-lived, such that the 'pure' blocked/unblocked phases dominate the transport. Formulating this precisely requires further insight into the switching process, and we reserve this discussion for the following section: here, we shall simply formulate the arguments assuming transients to be 'sufficiently fast'.
\\
In this case, a prediction for the total current follows for each of the three different scenarios introduced above:
\\
\begin{itemize}
\item
LD$\to$HD$^*$: in this scenario, the single-phase current is $J_{LD}=\alpha_A(1-\alpha_A)$. Using Eq. (\ref{eq:tauHD}) for the 'blockage time, $\tau\bl$, and introducing $\tau\unbl$ for the 'unblocked' time during which there is free flow, we therefore have
$J_{LD \to HD^*} = J_{LD} \frac{\tau\unbl}{\tau\bl+\tau\unbl}$.
This is essentially a weighted average of the current, since we expect to have current $J_{LD}$ during the time interval $\tau\unbl$ and zero current the rest of the time, with $\tau\unbl = \frac{n_A}{J_{LD}}$. Thus we have
\begin{equation}
  \label{eq:LD-HD}
  J_{LD\to HD^*} = \frac{\alpha_A \, \beta_B}{\frac{\alpha_A\beta_B}{\alpha_A(1-\alpha_A)}+\alpha_B}.
\end{equation}

\item MC$\to$HD$^*$: following the same approach as above, we obtain 
\begin{equation}
  J_{MC\to HD^*}  = \frac{1}{4} \frac{\frac{4\alpha_A}{\alpha_B}}{\frac{4\alpha_A}{\alpha_B}+\frac{1}{\beta_B}}
  \ .
\end{equation}

\item HD$\to$HD$^*$: in this case, we obtain: 
\begin{equation}
  \label{eq:hd-hd}
  J_{HD \to HD^*} = \frac{\alpha_A\,\beta_B}{\frac{\alpha_A\,\beta_B}{\beta_A(1-\beta_A)}+\alpha_B}\ .
\end{equation}

\end{itemize}

Notice that in Eq.~(\ref{eq:hd-hd}), as both exit rates $\beta_A$ and $\beta_B$ vanish, $J_{HD-HD} \approx \beta_{eff}$, in agreement with  the conventional TASEP current in HD:
\begin{equation}
  \label{eq:jhdapprox}
  J_{HD} = \beta(1-\beta) \approx \beta
  \ .
\end{equation}

Predictions from this 'intermittent' mean-field (iMF) theory for the current are shown in Figs. \ref{fig:LD-HD:J}, \ref{fig:MC-HD:J} and \ref{fig:HD-HD:J} (black circles), superposed onto the standard mean-field results (blue circles) as well as data from simulations (red circles).
The latter correspond to Gillespie simulations which were run until $10^5$ slow B particles had entered the system, thus leading to roughly that number of blockage events at the exit. 
The insets show the relative error of the current for both the MF (blue circles) and iMF approaches (black circles). They clearly show that in the LD$\to$HD$^{*}$ and MC$\to$HD$^{*}$ scenarios  the iMF approach performs better than MF in the limit of very small values of $\chi_B$. For intermediate $\chi_B$ values, both approaches are comparable. For larger values of $\chi_B$, the MF performs better than the iMF: this is as expected, since for larger values of $\chi_B$, where the proportions of A and B particles are comparable, there should be no intermittent behaviour. A special case is the HD$\to$HD$^{*}$ scenario, for which iMF and MF perform indistinguishably well for both small and intermediate values of $\chi_B$. This is because in the HD$\to$HD$^{*}$ scenario both $\beta_A$ and $\beta_B$ are very similar, and therefore the difference in the MF and iMF expressions is very small, as pointed out in Eq.~\ref{eq:jhdapprox}.

In Figs.~\ref{fig:HD-LD:rho}, \ref{fig:HD-MC:rho} and \ref{fig:HD-HD:rho} the insets show the comparison of the average density predicted by the MF (dashed line), iMF (dotted line) and the numerically obtained density profile (solid line) for a fixed value of $\chi_B$. Parameters are chosen to be well within the intermittent regime, and the average density in the iMF has been attributed by equating the current to the mean-field expression $\rho(1-\rho)$ and solving for $\rho$.
Comparing the different approximations for the density and the corresponding numerical simulations thus mirrors what is observed for the currents.
\\

The iMF approach therefore successfully takes over from the standard MF description when B particles are very slow to leave, as far as they are remain sparse compared to A particles. The limitations of the approach lie in assuming that the current of $A$ particles is stationary.
\\

 In reality, we know that whenever a slow B particle frees the exit, a highly non-stationary process will ensue, during which the density profile relax from a totally jammed state at the exit to the stationary density profile corresponding to a 'pure' flow of $A$ particles in the appropriate phase (LD, MC or HD).
The iMF approach as outlined here therefore assumes that these non-stationary phases remain sufficiently short so that they do not affect the time-averaged density. This assumption must fail when blockages become too frequent for this to be true, or even too frequent for the density to return to its stationary density before the next blockage: clearly, this is the reason why iMF predictions perform poorly as the fraction of B particles becomes significant.

\section{Mechanism of transients and validity of iMF predictions\label{sec:mechanisms}}

\begin{figure*}
  \centering
    \includegraphics[width=0.4\textwidth]{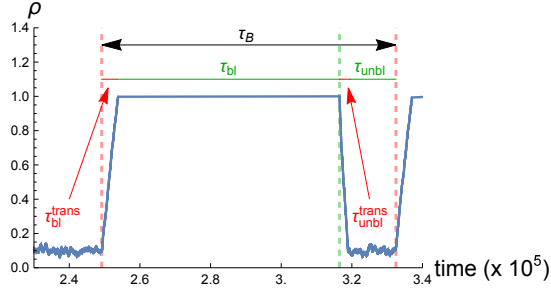} 
  \caption{\label{fig:sketch:timescales}
    Schematic illustration of the timescales in an iMF scenario. The density and the current in the system stochastically switch between two stationary phases, one during which the exit site is blocked by a B particle, the other one is the time during which the current flows freely~: the corresponding timescales are $\tau\bl$ and $\tau\unbl$, respectively. The time for the transient towards the blocked state is noted $\tau\bl\trans$, and $\tau\unbl\trans$ denotes the equivalent time for the transient towards the unblocked state. Also shown is the timescale $\tau_B$, which represents the (average) time between two blocking events. For iMF arguments to be valid, the transients must be short-lived, so that the stationary states characterise the system essentially all of the time. The dashed vertical lines indicate the times of blocking/unblocking events: a B particle arrives at the exit site at the green lines, and leaves the lattice at the red lines.
  }
\end{figure*}

The question has been raised above to which extent the iMF arguments can be expected to hold, as it assumes switching between essentially two states, a 'blocked'  and an 'unblocked' one, the last one corresponding to a  free-flowing 'pure' state of only $A$ particles. These two states are considered as stationary, whereby transitory states, which are schematically shown in Fig.~\ref{fig:sketch:timescales}, are implicitly neglected. We denote the timescales for the stationary states as  $\tau\bl$ (the average time interval during which we have a stationary blocked state) and $\tau\unbl$ (the average duration of the stationary unblocked, free-flowing state). Transient times are referred to as  $\tau\unbl\trans$ (for the 'unblocking' process, i.e. the switching from the blocked to the free-flowing stationary state), and  $\tau\unbl\trans$ (for the reverse transient).

For iMF to make correct predictions, the stationary states must therefore be sufficiently long-lived to dominate the averages, i.e. we must have
\begin{equation}
  \label{eq:iMF:condition}
  \tau\bl\trans + \tau\unbl\trans \ll \tau_B
  \ ,
\end{equation}\label{eq:cond1}
where $\tau_B$ is the (average) time between two successive arrivals of blocking B particles at the exit site. 
Note that $\tau_B$ thus comprises both stationary and transient times (see Fig. \ref{fig:sketch:timescales}), i.e. we have
\begin{equation}
  \tau_B=\tau\bl\trans+\tau\bl+\tau\unbl\trans+\tau\unbl
  \ .
\end{equation}

The approach here is to view the 'blocking' event, where a B particle arrives at the exit site, and the 'unblocking' event, when the B particle eventually leaves the system, as abrupt changes of boundary conditions. It is known~\cite{santen:2002} that such changes can be conveniently described as setting in motion a domain wall, i.e. a singularity  which separates two zones of different densities. Due to the imbalance of currents on both sides of this discontinuity, the position of the domain wall evolves. 
Specifically, there is a straightforward expression for the speed at which such a domain wall evolves, which is  \cite{kolomeisky:1998}
\begin{equation}
  \label{eq:dw:v}
  V_{DW} = \frac{J_- - J_+}{\rho_--\rho_+} 
\end{equation}
where '-' and '+' refer to the phases to the left and to the right of the domain wall, respectively.
We now exploit this relation for the specific changes following the blocking and unblocking events, in order to estimate the associated transient times $\tau\bl\trans$ and $\tau\unbl\trans$. This requires fixing the specific situation one is interested in; here we shall discuss the $LD \to HD^*$ scenario, where a B particle blocking the exit pushes the system from an LD state into a HD-like state.

\begin{figure*}
  \centering
  \subfloat[]{\label{fig:mechanisms:a}
    \includegraphics[width=0.45\textwidth]{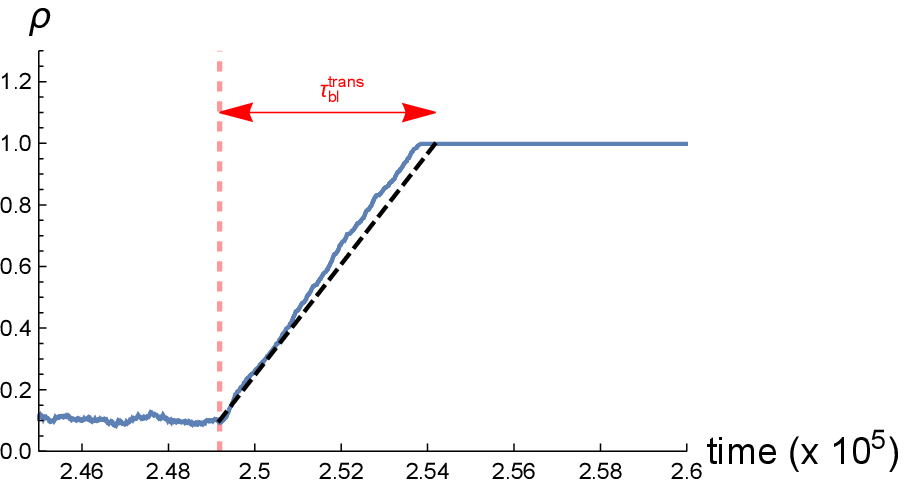} 
  }
  \subfloat[]{\label{fig:mechanisms:b}
    \includegraphics[width=0.45\textwidth]{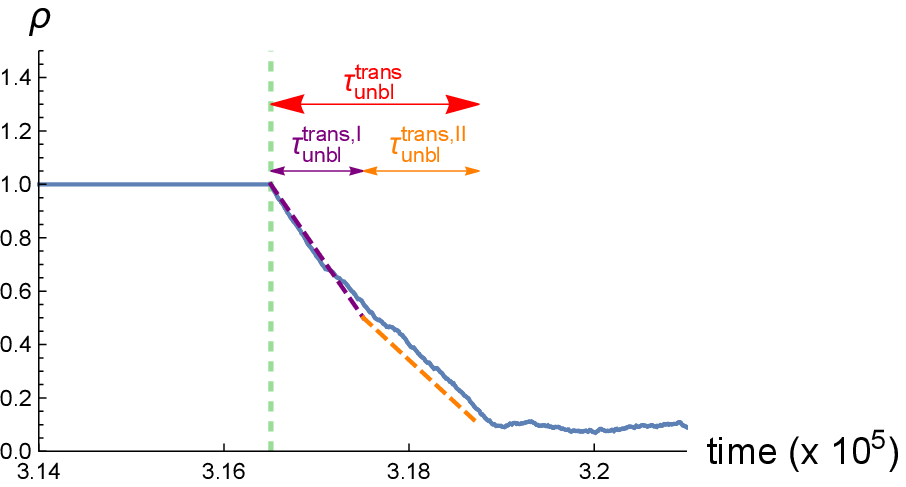} 
  }
  \caption{\label{fig:mechanisms}
    Numerical illustration of the domain wall propagation mechanisms involved in the transients. The average density (or, equivalently, the total number of particles in the segment) directly reflects the position of the domain wall.
    (a) After a blocking event, as a B particle arrives at the exit site, a domain wall propagates upstream.
    (b) After the following unblocking event, as the B particle frees the exit, two successive domain walls cross the segment to take it to its new stationary state.
    All simulations have been performed with the following parameters: $\alpha_A=0.1$, $\beta_A=1$, $\alpha_B=1.e-5$, $\beta_B=1.e-5$. The transient times predicted by Eqs.~(\ref{eq:dw:blocking}
), ~(\ref{eq:dw:unblocking}) and ~(\ref{eq:dw:unblocking2}) are also shown (double arrows). Based on these, dashed lines (black, purple and orange) are derived, invoking the density difference $\rho_+ - \rho_-$ within those time intervals.
  }
\end{figure*}

First, we estimate $\tau\bl\trans$. We  thus consider the situation where there is a free-flowing LD phase, and a slow-to-leave B particle arrives at the exit site, where it remains for a significant time to come. 
This effectively sets the exit rate for the particles in the segment to 0, and the sites close to the exit will progressively fill up to saturation ($\rho_+=1$). The resulting domain wall therefore separates a free-flowing $LD$ phase, with density $\rho_-=\alpha_A$ and current $J_- = \alpha_A \, (1-\alpha_A)$, to its left, and a fully blocked region with $\rho_+=1$ and $J_+=0$, to its right.
Applying relation (\ref{eq:dw:v}) yields a domain wall velocity of
\begin{equation}
  \label{eq:blocking:Vdw}
  V_\mathrm{dw,blocking} =  \frac{\alpha_A \, (1-\alpha_A) - 0}{\alpha_A - 1} =- \alpha_A
  \ ,
\end{equation}
i.e. the domain wall propagates towards the entry site, as expected. From this we estimate the transient time for the system of $L$ sites to become fully blocked as
\begin{equation}
  \label{eq:dw:blocking}
  \tau\bl\trans \approx \frac{L}{\alpha_A}
  \ .
\end{equation}
This implicitly assumes that the entire segment saturates before the B particle eventually leaves the exit site, and $\tau\bl\trans$ therefore constitutes an upper bound for the transient time. A numerical observation of this process is illustrated in Fig.~\ref{fig:mechanisms:a}, which quantitatively confirms the argument.

Next, we estimate $\tau\unbl\trans$. To this end we consider the situation when the blocking B particle ultimately frees the exit site. This again amounts to a modification of the exit rate for the particles in the segment, which now returns to $\beta_A$,  and thus again sets off a domain wall which propagates upstream to the entry site. 
This process establishes a new bulk density, which is determined by the new exit rate $\beta_A$, but still based upon a zero input rate, since the entrance remains blocked right until it is freed up by the arrival of the the domain wall.
Therefore, once the domain wall arrives at the first site of the lattice, 
 a {\it second} domain wall follows, triggered at the entrance this time, and which needs to propagate back to the exit before the definite steady state is reached. We denote (transient) time scales for these two stages as $\tau\unbl\trans{}^{,I}$ and $\tau\unbl\trans{}^{,II}$, respectively.

Both phases of this two-stage process can be treated by adapting the argument given above to calculate the domain wall velocity. However, the specific values of the successive densities in the intermediate phases depend not only on the fact that we are dealing with a LD $\to$ HD$^*$ scenario, but also on the actual values of the boundary rates $\alpha_{A,B}$ and $\beta_{A,B}$. These calculations are a little lengthy but straightforward, and they are therefore confined to Appendix \ref{app}. Here we simply refer to Fig.~\ref{fig:mechanisms:b}, which shows that the segment-averaged particle density indeed evolves as predicted by the two-stage picture.

The data presented so far illustrates that the transient mechanisms can indeed be understood in terms of propagating domain walls. However, ideally one would like to be able to formulate a criterion capturing to which extent the iMF approach may be expected to be successful. To this end, we recall that the main result of our calculations is an expression for the timescales required to unblock the system, which we summarise, using the results from Appendix \ref{app}, as:
\begin{equation}
    \label{eq:dw:unblocking}
  \tau\unbl\trans{}^{,I} =
  \left\{
  \begin{array}{ccccc}
    \frac{L}{1-\beta_A} & (\alpha_A<\beta_A<1/2) \\
    2L                  & (\alpha_A<1/2<\beta_A)
  \end{array}
  \right.
  \ ,
\end{equation}
and
\begin{equation}
 \label{eq:dw:unblocking2}
  \tau\unbl\trans{}^{,II} =
  \left\{
  \begin{array}{ccccc}
    \frac{L}{\beta_A-\alpha_A} & (\alpha_A<\beta_A<1/2) \\
    \frac{L}{\frac{1}{2}-\alpha_A}       & (\alpha_A<1/2<\beta_A)
  \end{array}
  \right.
  \ ,
\end{equation}
all of which refer to the scenario $LD \to HD^*$.
With these predictions, as well as equation (\ref{eq:dw:blocking}) for the blocking process, one can thus expect iMF predictions to work when the total transient time $\tau\trans \ll \tau_B$.
\\

We also recall that $\tau_B$ is the (average) time lapse between two blocking events. It can be estimated based on the entry current, requiring that a single B particle enters the system. For the $LD \to HD^*$ scenario this reads
\begin{equation}
  1 = J_{\mbox{in}} \, \tau_B = \alpha_B \, (1-\rho_1) \times \tau\unbl + 0 \times \tau\bl
  \ ,
\end{equation}
where $\rho_1$ is the average density on the first site of the segment during the unblocked phase. Using a self-consistent argument, we assume the LD expression for the density during this unblocked phase, $\rho_1 = \alpha_A$, which yields
\begin{equation}
  1 = \alpha_B \, (1-\alpha_A) \, \ \tau\unbl
  \ ,
\end{equation}
and we can thus estimate
\begin{equation}
  \tau\unbl  \approx \frac{1}{\alpha_B \, (1-\alpha_A)}
  \ .
\end{equation}
We now obtain $\tau_B$ by adding to this the time it takes for a blocking $B$ particle to exit, which is simply given by its exit rate as $\tau\bl \approx 1/\beta_B$, so that
\begin{equation}
  \label{eq:tauB}
\tau_B \approx \frac{1}{\beta_B} + \frac{1}{\alpha_B \, (1-\alpha_A) }
\ .
\end{equation}

\begin{figure*}
  \centering
  \subfloat[]{\label{fig:iMF:criterion:a}
    \includegraphics[width=0.3\textwidth]{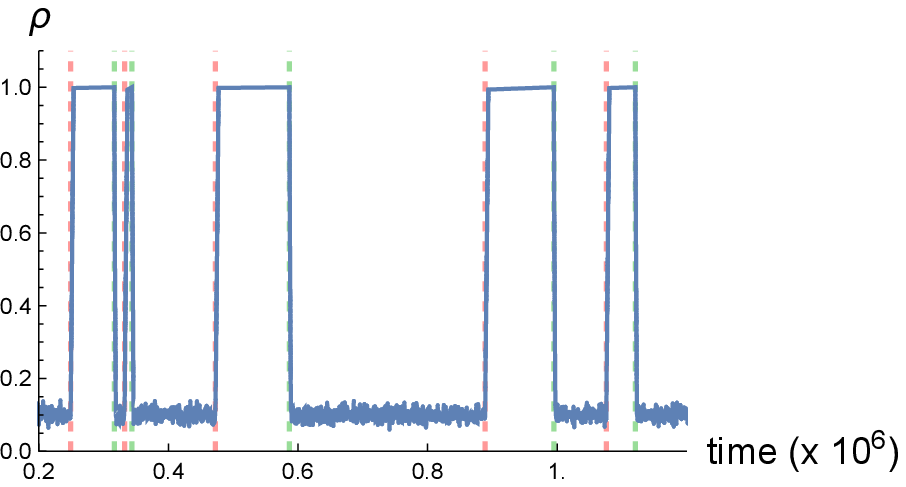}
  }
  \subfloat[]{\label{fig:iMF:criterion:b}
    \includegraphics[width=0.3\textwidth]{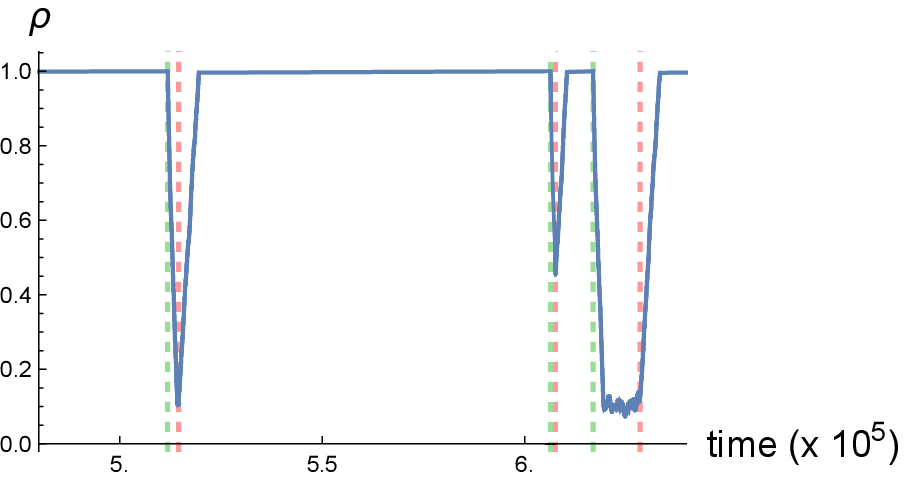}
  }
  \subfloat[]{\label{fig:iMF:criterion:c}
    \includegraphics[width=0.3\textwidth]{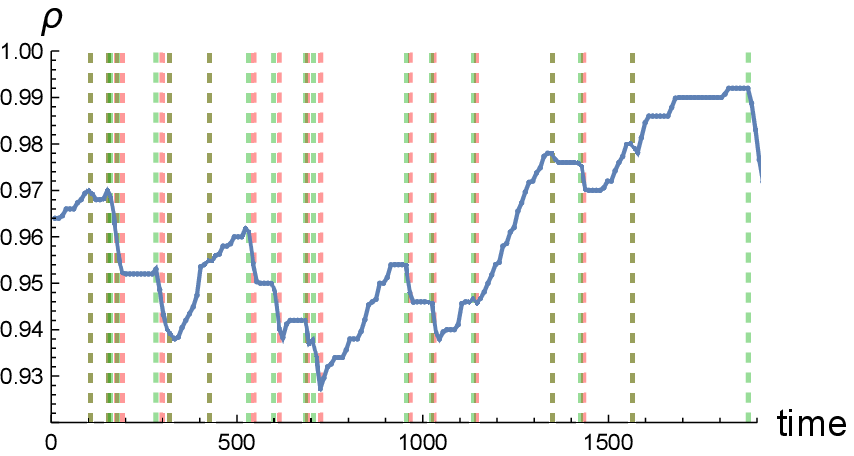}
    }
     \caption{\label{fig:iMF:criterion}
      Predictions based on iMF theory require the stationary states to dominate the system. Simulation data is shown for the average density in the segment as a function of time. All plots sharing the parameters 
      $\alpha_A=0.1$, $\beta_A=1$, but rates for $B$ particles vary as follows:
      (a) $\alpha_B=1.e-5$, $\beta_B=1.e-5$
      (b) $\alpha_B=1e-4$, $\beta_B=1e-5$
      (c) $\alpha_B=0.05$, $\beta_B=0.01$
      Whereas the plot (a) corresponds to a case where iMF is expected to apply, this is not the case for the other examples. In (b) the B particles are not sufficiently rare, so that the free-flowing regime cannot develop. In (c) they are also not sufficiently slow-to-leave, so that no proper system-wide blockage can be observed. The lattice size used forthese simulations was $L=500$.
    }
\end{figure*}

Based on this expression, as well as the condition (\ref{eq:iMF:condition}) and the expressions for the transitory times Eqs.~(\ref{eq:dw:blocking}) and (\ref{eq:dw:unblocking}), a criterion can thus be formulated for iMF to be valid. This has to be done separately for cases (i) and (ii): iMF works when the B particles are both 'sufficiently slow' and''sufficiently sparse', in a system size dependent sense. Interestingly, the criterion is seen to be more restrictive close to the phase boundaries. The way in which iMF succeeds or fails is illustrated in Fig.~\ref{fig:iMF:criterion}. Please refer to Appendix \ref{app} for a quantitative derivation of the corresponding criteria.

\section{Discussion}

In this paper we have analysed an extension of the TASEP model where we consider two different types of particles. No overtaking is allowed,  in contrast to most multi-species TASEP previously introduced in the literature~\cite{crampe,arita-mallick,derrida:1997}. Hence, our model is similar to the one introduced in~\cite{bottero}, which describes different types of molecular motors moving along microtubules. At the same time our model is complementary, as we focus on an entirely different scenario, considering that particle species have the same bulk hopping rate but differ both in their entry and exit rates with which enter and leave the lattice, respectively.

We have shown that a standard mean-field theory can be formulated by mapping the two-species TASEP model onto an effective single-species TASEP, via appropriately defined effective entry/exit rates.  A comprehensive phase diagram can be established based on the ensemble of all entry/exit rates for both particle species, according to which different scenarios arise. In these several, but not necessarily all, possible TASEP phases are present. For example, when the exit rates of both particles are inferior to half of the bulk hopping rate, then there cannot be a maximum current (MC) phase. Comparison to stochastic simulations has shown that this approach yields excellent results for the current and for the density profile along the lattice  as long as the entry and exit rates of the different types of particles are of the same order of magnitude.

The mapping onto an effective single-species mode fails, however, when one type of particles is much rarer and has a much slower exit rate than the other one. We have shown that the origin of this discrepancy lies in the emergence of intermittent dynamics caused by temporary blockages of the exit by the slow-to-leave particles.
The key to the intermittent regime is to analyse the transport process in terms of traffic jams which form close to the exit, whenever a slow-to-leave particle arrives at the last lattice site. This picture constitutes a valid representation, based on which we have introduced a modified mean-field approach. This 'intermittent mean-field' (iMF) description takes into account the intermittent behaviour, and we have shown that it provides good predictions when compared to simulations.

\begin{figure*}
    \includegraphics[width=1\textwidth]{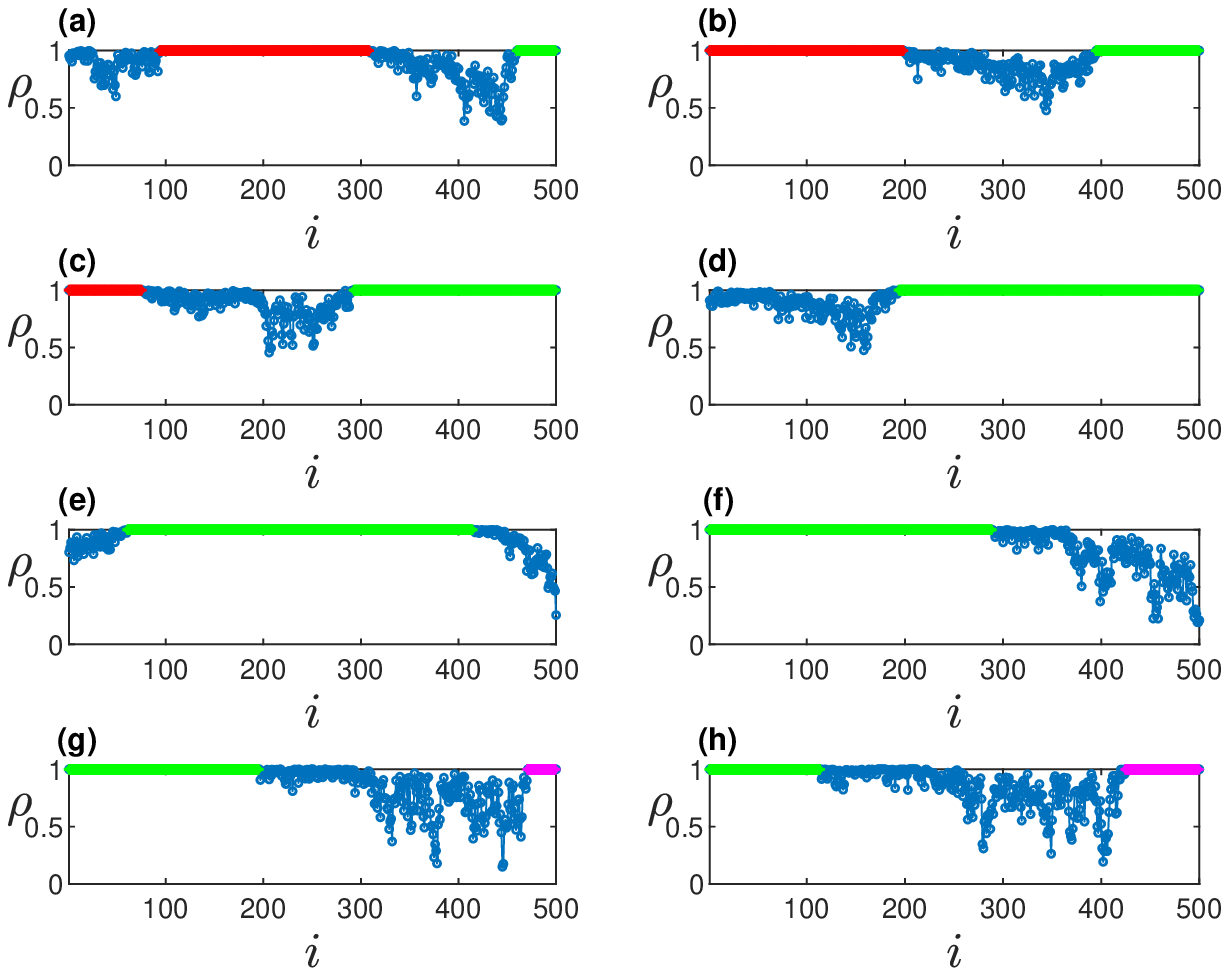}
  \caption{
    \label{fig:intermittency:timeseries2}
    Time evolution of the density profiles for the $MC\to HD^{*}$ scenario, showing the evolution of 3 temporary blockages of the exit site and their evolution. Panels (a) to (h) represent the succession of instantaneous-like density profiles along the same simulation. Parameters are $\alpha_A = 1$; $\alpha_B = 0.05$; $\beta_A = 1$; $\beta_B = 0.003$. Each snapshot has been calculated by averaging over 1,300 Gillespie iterations (each iteration corresponding to one reaction occurring in the system, i.e., the movement of one particle on the lattice). The number of Gillespie iterations separating successive snapshots is equal to 7,800. The blockages created by B particles are highlighted in red, green and pink, in order of appearance. Successive snapshots are ordered in time but do not correspond separated by identical time intervals: they have been selected to illustrate the essence of the process with only a few snapshots.
 }
\end{figure*}

Dynamic features of traffic jams reveal additional questions in their own right. We have shown that viewing the 'jammed' region as being delimited by two abrupt changes, and treating these discontinuities in the 'domain wall' picture  \cite{kolomeisky:1998,santen:2002}, leads to a valid and useful description. In particular, this has allowed us to establish the conditions under which iMF is expected to hold:  the slow-to-leave particles must be  both sufficiently slow and sufficiently rare, in a sense which depends on the system size and on the proximity to the phase boundaries of the fast particles.

Closer inspection shows, however, that there are yet more subtle features to the dynamics. For example, since the size of a jammed region evolves due to the motion of its delimiting domain walls, a jam may reach the entrance of the lattice, where it will then dissolve by shedding its remaining particles into the system. But it may also cease to exist before reaching the boundary, which again points to the importance of system size. Further processes can arise: for example, we have observed the traffic jam to split into two or more jams, through a mechanism where holes penetrate into the jammed region from its downstream side (see Fig.~\ref{fig:intermittency:timeseries} (c)). 

According to the parameter regimes, a full description of the transport process would thus have to account for the simultaneous presence of several moving traffic jams (as is indeed captured by the snapshot in Fig.~\ref{fig:intermittency:timeseries2}). These evolve and move, but also interact, presumably through processes reminiscent of fission and coalescence.
The interplay of all these non-stationary processes promises rich behaviour indeed, which will again depend on system size the system size.

Another logical next step would be to analyse a model in which two species are distinguished not only by their entry and exit rates, but also by their bulk hopping rates. This would effectively combine our model with that by Bottero et al.~\cite{bottero}. The phenomenology in this six-parameter model may be expected to be rich. Fully differentiating all microscopic rates for the two particle species would then provide a solid starting point for exploring applications, such as translation by ribosomes from mRNA which may or may not have bound RAC/NAC proteins \cite{racnac}, or  transport of different types of ions through membrane channels~\cite{kolomeisky}.

\section*{Acknowledgments}
This work has been partly funded by SULSA (Scottish Universities Life Science Alliance). We would also like to thank the Visitors Scholar Programme from the University of Aberdeen for funding N. Kern visits to Aberdeen, during which a substantial part of this work was developed.

\appendix

\section{Transitory time scales for the $LD \to HD^*$ scenario\label{app}}

In the main text, section \ref{sec:mechanisms},  we have established that the transients can be understood as domain walls propagating through the system. With the understanding of these processes, we can then establish the timescales of these transients. It is these which ultimately lead to conditions for the applicability of the intermittent mean-field approach. Here we complete these arguments presented for the $LD \to HD^*$  scenario.

\subsection{Blocking transient}

For a first transient process, which takes the system from an unblocked to a blocked state, the velocity of the domain wall has been shown to be given by Eq.~(\ref{eq:blocking:Vdw}), from which we have deduced an associated timescale of (see Eq.~\ref{eq:dw:blocking})
\begin{equation}
  \label{eq:dw:blocking:recall}
  \tau\bl\trans \approx \frac{L}{\alpha_A}
\end{equation}
for a segment of $L$ sites, assuming that the entire segment ends up being jammed.

We now need to establish the equivalent for the unblocking process. Before doing so, however, it is useful to formally state conditions which we have already made when attempting to apply iMF arguments to the entire segment.

First of all, it is intuitively clear that B particles must be 'sparse' in some sense. We can formalise this by requiring that the fraction of B particles, which is given by Eq.~(\ref{eq:chiAB:def}), must be small enough so that the average spacing between them exceeds the lattice length.
Therefore
\begin{equation}
\chi_B = \frac{\alpha_B}{\alpha_A+\alpha_B}  \ll \frac{1}{L}
\end{equation}
is required. This condition can be simplified to
\begin{equation}
  \label{eq:iMF:condition:sparse}
  \alpha_B \ll \frac{\alpha_A}{L}
  \ ,
\end{equation}
which means that the criterion for 'sparseness' is system size dependent.

Second, we can state that the transient must last for long enough so the domain wall can move right up to the entrance before the exit is unblocked. Using Eq.~(\ref{eq:dw:blocking:recall}) this yields
\begin{equation}
  \tau\bl\trans = \frac{L}{\alpha_A} \ll \frac{1}{\beta_B}
  \ ,
\end{equation}
since $1/\beta_B$ is the (average) time for a B particle to leave the exit site. Writing this as
\begin{equation}
  \label{eq:slow}
  \beta_B \ll \frac{\alpha_A}{L}
\end{equation}
makes it clear that this requires B particles to be 'sufficiently slow to leave', again in a sense dependent on system size.

Note that, since we are interested in the $LD \to HD^*$ scenario here, we must have $\alpha_A<\beta_A$ for the pure A phase to be in HD, and therefore Eq.~(\ref{eq:slow}) implies
\begin{equation}
  \label{eq:sparse}
  \beta_B \ll \frac{\beta_A}{L}
  \ ,
\end{equation}
which thus provides a condition similar to that for 'sparseness'.

\subsection{Unblocking transient}

To estimate the unblocking transient time $\tau\unbl\trans$, we need to consider two stages, each one corresponding to the propagation of a domain wall, first upstream (stage I) and then downstream (stage II). We now derive the corresponding time scales, which have been stated without proof in the main text.

In stage I, no particles enter from the left ($J_-=0$) as the jam is dense ($\rho_-=1$) right up to the entrance site. The phase to the right of the domain wall, however, carries particles to the exit. This zone can be visualised as receiving particles from the jam (in-rate $\alpha_+=1$), and having an out-rate $\beta_+=\beta_A$ once the B particle blocking the exit site is gone. This second zone therefore corresponds either to a $HD$ zone  (case (i), if $\beta_A<1/2)$, or to an MC zone (case (ii), if $\beta_A>1/2$). Recall that we already have the condition $\alpha_A<\beta_A$, as we are dealing with an $LD \to HD^*$ scenario.  We can therefore distinguish two cases:
\\

\paragraph{Unblocking transient in case (i)  $\alpha_A < \beta_A < 1/2$ :}

In this case, at the beginning of stage I, the segment at the exit is in an HD phase. We thus have $\rho_-=1$ and  $J_-=0$, as well as $\rho_+ = 1-\beta_A$ and $J_+=\beta_A \, (1-\beta_A$). The domain wall velocity is therefore
\begin{equation}
  V_{DW,I} = \frac{0 - \beta_A \, (1-\beta_A)}{1-(1-\beta_A)}  = - (1-\beta_A)
  \qquad\mbox{(case i)}
\ .\end{equation}
This is negative, as expected, as the DW moves upstream towards the entrance.
\\

Therefore the time $\tau\unbl\trans{}^{,I}$ required for this (transient) DW to reach the entrance is
\begin{equation}
\tau\unbl\trans{}^{,I} = \frac{L}{\left|V_{DW,I}\right|} = \frac{L}{1-\beta_A}
\qquad\mbox{(case i)}
\ .
\end{equation}

After this time, on average, the entire system is in an HD phase at density $1-\beta_A > 1/2$.
\\

Now stage II begins, as the in-rate $\alpha_A$ acts at the freshly unblocked entrance, thus initiating a new DW which travels travel downstream.
Therefore, the newly created zone at then entrance is governed by an in-rate $\alpha_-=\alpha_A < 1/2$, according to case (i), whereas the rate for exiting to the right of the DW, a zone of density $\rho_+ = 1-\beta_A > 1/2$, is $\beta_- = 1-\rho_I = \beta_A < 1/2$, as established throughout the lattice by the previous stage I.
Since we also know $\alpha_A<\beta_A$, from case (i), we thus know that the initial segment is in an LD phase. Therefore $\rho_-=\alpha_-=\alpha_A$ and $J_- = \alpha_A \, (1-\alpha_A)$.
\\

The zone to right of the DW being in HD, at density $\rho_+=\rho_I = 1-\beta_A > 1/2$, we can thus deduce the velocity of this second $DW_{II}$ as
\begin{equation}
  V_{DW,II} = \frac{\alpha_A (1-\alpha_A) - \beta_A \, (1-\beta_A)}{\alpha_A - (1-\beta_A)}
  \qquad\mbox{(case i)}
\end{equation}
which, after simplification, leads to
\begin{equation}
  V_{DW,II} = \beta_A-\alpha_A
  \qquad\mbox{(case i)}
    \ .
\end{equation}
This is indeed positive, as expected, so we can afirm that this domain wall will cross the system and instore a new density.
\\

Based on this domain wall velocity, the second timescale for returning to a stationary state is
\begin{equation}
  \tau\unbl\trans{}^{,II} = \frac{L}{\beta_A-\alpha_A} \qquad\mbox{(case i)}
  \ .
\end{equation}

All in all, for case (i), the condition of validity for iMF stateded in Eq.~(\ref{eq:iMF:condition}), therefore reads
\begin{equation}
  \label{eq:iMF:condition:i}
  \frac{L}{\alpha_A} + \frac{L}{1-\beta_A} + \frac{L}{\beta_A-\alpha_A} \ll \tau_B
\qquad\mbox{(case i)}\ .
\end{equation}
Before exploiting this condition further, we now perform the equivalent analysis for case (ii).
\\

\paragraph{Unblocking transient in case (ii)  $\alpha_A < 1/2 < \beta_A$ :}

In this case, as soon as the blocking B particle leaves the exit site, a depletion zone opens up into which particles enter from the blocked zone with rate 1, and from which they exit at rate $\beta_A$. Since $\beta_A>1/2$, we are thus dealing with an MC phase, and we have $\rho_+=1/2$ and $J_+=1/4$. From this we have the domain wall velocity
\begin{equation}
  V_{DW,I} = \frac{0 - 1/4}{1-1/2}  = -1/2
  \qquad\mbox{(case ii)}
\ ,\end{equation}
and the corresponding timescale is 
\begin{equation}
\tau\unbl\trans{}^{,I} = \frac{L}{\left|V_{DW,I}\right|} = 2 \, L
\qquad\mbox{(case ii)}
\ .
\end{equation}

In stage II, after the MC phase has filled the segment, a zone develops at the entrance into which particles attempt to enter at rate $\alpha_A$, and from which they leave at rate $1-1/2 = 1/2$. Consequently, this is an LD zone, and we thus have $\rho_-=\alpha_A$ and $J_- = \alpha_A \, (1-\alpha_A)$. From this,
\begin{equation}
  V_{DW,II} = \frac{\alpha_A (1-\alpha_A) - 1/4}{\alpha_A - 1/2}
  = \frac{1}{2}-\alpha_A
  \qquad\mbox{(case ii)}
\end{equation}
and therefore 
\begin{equation}
  \tau\unbl\trans{}^{,II} = \frac{L}{1/2-\alpha_A} \qquad\mbox{(case ii)}
  \ .
\end{equation}

Condition~(\ref{eq:iMF:condition}) for iMF to hold becomes therefore
\begin{equation}
  \label{eq:iMF:condition:ii}
\frac{L}{\alpha_A} + 2 \, L + \frac{L}{1/2-\alpha_A} \ll \tau_B
\qquad\mbox{(case ii)}
\ .
\end{equation}

\subsection{Conditions of validity for iMF}
The conditions for iMF to be valid have been stated, separately for cases (i) and (ii), in Eqs. (\ref{eq:iMF:condition:i}) and (\ref{eq:iMF:condition:ii}). They are based on establishing orders of magnitude, and we can therefore simplify them further by involing the following arguments:
\\

As a first observation, the expression  Eq.~(\ref{eq:tauB}) for $\tau_B$ can be adapted as
\begin{equation}
  \label{eq:tauB:approx}
  \tau_B
  \approx \frac{1}{\beta_B} + \frac{1}{\alpha_B \, (1-\alpha_A) }
  \simeq
  \frac{1}{\beta_B} + \frac{1}{\alpha_B}
  \ ,
\end{equation}
which is valid since $\alpha_A < 1/2$: the neglected factor is therefore of order unity, and does not change our comparison of orders of magnitudes.
To simplify further, we remark that in terms of orders of magnitude this is essentially equivalent to
\begin{equation}
  \label{eq:tauB:meaningful}
  \tau_B \simeq \frac{1}{\mbox{min}(\alpha_B,\beta_B)}
  \,
\end{equation}
since the term with the smaller denominator dominates $\tau_B$. This shows that $\tau_B$ characterises the larger one of the timescales associated to the entrance and exit rates of B particles. We will use this expression in the following.
\\

Second, we observe that both conditions of validity (Eq.~(\ref{eq:iMF:condition:i}) for case (i) and Eq.~(\ref{eq:iMF:condition:ii}) for case (ii), respectively), are based on a sum of three positive terms being negligible compared to $\tau_B$. The timescale separation expressed by these two requirements can therefore only occur if each of these terms is small compared to the right-hand side. We are therefore dealing with {\it three} conditions. For case (i), for example, we have
\begin{equation}
  \label{eq:iMF:3conditions:i}
  \frac{L}{\alpha_A} \ll \tau_B
  \ , \ \ 
  \frac{L}{1-\beta_A} \ll \tau_B
  \ \ \mbox{and} \ \ 
 \frac{L}{\beta_A-\alpha_A} \ll \tau_B
 \qquad\mbox{(case i)}
\end{equation}
and {\it all} of these conditions must be satisfied. For case (ii) they read
\begin{equation}
  \label{eq:iMF:3conditions:ii}
  \frac{L}{\alpha_A} \ll \tau_B
  \ , \ \ 
  2 \, L \ll \tau_B
  \ \ \mbox{and} \ \ 
  \frac{L}{1/2-\alpha_A} \ll \tau_B
 \qquad\mbox{(case ii)} 
\end{equation}

Third, note that the first one of these conditions is common to both cases (i) and (ii), $\alpha_A \gg L\tau_B$. It has in fact already been evaluated above, leading to the requirement of 'slowness', stated in Eq.~(\ref{eq:slow}), and therefore there is no further consideration to be had here.
\\

All in all are thus left with last two requirements of Eq.~(\ref{eq:iMF:3conditions:i}) for  case (i),  and of Eq.~(\ref{eq:iMF:3conditions:ii}) for  case (ii): we will analyse these now.
\\

\paragraph{Conditions of validity for iMF in case (i) $\alpha_A < \beta_A < 1/2$:}

Considering the remaining conditions for case (i), and using Eq.~(\ref{eq:iMF:3conditions:i}), a sufficient condition for iMF to hold is that we have
\begin{equation}
  \frac{L}{1-\beta_A} \ll \tau_B  
  \ \ \mbox{as well as} \ \ 
  \frac{L}{\beta_A-\alpha_A} \ll \tau_B 
  \ .
\end{equation}

To analyse which condition is more stringent, there are two scenarios. If the terms on the left-hand sides are comparable, either of the conditions is sufficient. Otherwise, the larger one will lead to a more stringent condition. To this end, the first condition is more restrictive when $1-\beta_A < \beta_A-\alpha_A$, which can be expressed as $\beta_A > \frac{1+\alpha_A}{2}$.
Clearly this is never true, as $\beta_A<1/2$ and $\alpha_A>0$, and we must retain the second condition. Therefore, for iMF to work in case (i) we require, using Eq.~(\ref{eq:tauB:meaningful}) for $\tau_B$, 
\begin{equation}
  \frac{1}{\tau_B} \simeq \mbox{min}(\alpha_B,\beta_B) \ll \frac{\beta_A-\alpha_A}{L}
  \qquad \mbox{(case i)}
\end{equation}
which completes the requirement of 'sparseness', Eq.~(\ref{eq:sparse}), and of 'slowness', Eq.~(\ref{eq:slow}). Rather interestingly, this is a constraint on the rates for the slow-to-leave $B$ particles,  which becomes stronger for large systems, and which is particulary stringent in the vicinity of the phase boundary $\alpha_A=\beta_A$ of the 'pure' system.
\\

\paragraph{Conditions of validity for iMF in case (ii) $\alpha_A < 1/2 < \beta_A$:}
This case can be analysed in a similar fashion. From Eq.~(\ref{eq:iMF:condition:ii}), the extra conditions are
\begin{equation}
    2 L \ll \tau_B
    \ \ \mbox{as well as} \ \
    \frac{L}{\frac{1}{2}-\alpha_A} \ll \tau_B
  \ .
\end{equation}
The second condition is more stringent whenever $1/2-\alpha_A < 1/2 $, which is always true since $\alpha_A>0$. Therefore it is sufficient to require that
\begin{equation}
  \frac{1}{\tau_B} \simeq \mbox{min}(\alpha_B,\beta_B) \ll \frac{\frac{1}{2}-\alpha_A}{L}
    \ .
\end{equation}
for iMF to work in case (ii), in additon to 'sparseness'and 'slowness'. Again, this condition becomes impossible to meet for large enough systems, and it is particularly restrictive close to a phase boundary, here $\alpha_A=1/2$.

\bibliography{ttj}

\end{document}